\documentclass[preprint, 3p, 12pt]{elsarticle}
\usepackage{soul}
\usepackage{graphicx, graphics}
\usepackage{syntonly}
\usepackage{amssymb, amsmath, amsfonts}
\usepackage{subfigure}
\usepackage{multirow}
\usepackage{color}
\usepackage{bm}
\usepackage{hyperref}

\newcommand{\br}{\mathbf r}

\newcommand{\bx}{\mathbf x}
\newcommand{\bu}{\mathbf u}
\newcommand{\bD}{\mathbf D}
\newcommand{\bv}{\mathbf v}

\newcommand{\bF}{\mathbf F}
\newcommand{\bK}{\mathbf K}

\newcommand{\bS}{\mathbf P}

\newcommand{\ud}{\mathrm d}

\newcommand{\etal}{\textit{et al}}

\newcommand{\pfrac}[2]{\frac{\partial#1}{\partial #2}}
\newcommand{\cS}{\mathcal S}
\newcommand{\fF}{\mathcal F}

\newcommand{\fracp}[2]{\partial_{#2}{#1}}

\newcommand{\bsig}{\boldsymbol \sigma}

\newcommand{\tr}[1]{\mathrm{tr}({#1})}
\newcommand{\bvps}{\boldsymbol{\varepsilon}}

\renewcommand{\hl}{\textcolor{red}}
\newcommand{\tnabla}{\bar{\nabla}}
\newcommand{\tdelta}{\bar{\delta}}

\begin{document}
\begin{frontmatter}
\title{A nonlocal operator method for solving partial differential equations}
\author[BU]{Huilong Ren}
\author[HU,TU]{Xiaoying Zhuang\corref{cor1}}
\ead{zhuang@ikm.uni-hannover.de}
\author[KU0,KU]{Timon Rabczuk\corref{cor2}}
\ead{timon.rabczuk@tdt.edu.vn}
\cortext[cor1]{Corresponding author}
\cortext[cor2]{Corresponding author}
\address[KU0]{Division of Computational Mechanics, Ton Duc Thang University, Ho Chi Minh City, Viet Nam}
\address[KU]{Faculty of Civil Engineering, Ton Duc Thang University, Ho Chi Minh City, Viet Nam}
\address[BU]{Institute of Structural Mechanics, Bauhaus-University Weimar, 99423 Weimar, Germany}
\address[HU]{Institute of Continuum Mechanics, Leibniz University Hannover, Hannover, Germany}
\address[TU]{State Key Laboratory of Disaster Reduction in Civil Engineering, College of Civil Engineering,Tongji University, Shanghai 200092, China}
\begin{abstract}
We propose a nonlocal operator method for solving partial differential equations (PDEs). The nonlocal operator is derived from the Taylor series expansion of the unknown field, and can be regarded as the integral form ``equivalent'' to the differential form in the sense of nonlocal interaction. The variation of a nonlocal operator is similar to the derivative of shape function in meshless and finite element methods, thus circumvents difficulty in the calculation of shape function and its derivatives. {The nonlocal operator method is consistent with the variational principle and the weighted residual method, based on which the residual and the tangent stiffness matrix can be obtained with ease.} The nonlocal operator method is equipped with an hourglass energy functional to satisfy the linear consistency of the field. Higher order nonlocal operators and higher order hourglass energy functional are generalized. The functional based on the nonlocal operator converts the construction of residual and stiffness matrix into a series of matrix multiplications on the nonlocal operators. The nonlocal strong forms of different functionals can be obtained easily via support and dual-support, two basic concepts introduced in the paper. Several numerical examples are presented to validate the method.
\end{abstract}
\end{frontmatter}

%

\section{Introduction}\label{sec:introduction}

{In the field of solving PDEs numerically, many methods have been proposed, which include finite element method} \cite{zienkiewicz1977finite}, Smoothed Particle Hydrodynamics (SPH) \cite{lucy1977numerical}, Diffusive Element Method (DEM) \cite{nayroles1992generalizing}, Element-Free Galerkin (EFG) method \cite{Belytschko1994}, Reproducing Kernel Particle Method (RKPM) \cite{LiuJunZhang1995}, Partition of Unity Methods (PUM) \cite{babuvska1997partition}, Generalized Finite Element Method (GFEM) \cite{duarte2000generalized}, $hp$ clouds (HPC) \cite{duarte1996hp}, finite point method \cite{onate1996finite}, Generalized Finite Difference Method (GFDM) \cite{liszka1984interpolation}, the reproducing kernel collocation method \cite{aluru2000point,hu2011error}, Peridynamics \cite{Silling2007,Ren2015}, etc. Smoothed particle hydrodynamics developed by Lucy \cite{lucy1977numerical} in 1977 and Gingold and Monaghan \cite{gingold1977smoothed} estimates a function on support domain by the kernel approximation. In order to overcome the difficulty of discrete SPH of failing to reproduce constant fields, some other meshless methods have been proposed, two of the most famous methods are Moving Least Squares (MLS) and RKPM. The objective of MLS is to obtain an approximation based on the nodes in support, but with high accuracy and high order of completeness. Further developments are made in the Element-Free Galerkin method by Belytschko \etal \cite{Belytschko1994} and the RKPM proposed by Liu \etal \cite{LiuJunZhang1995} by increasing the order of completeness of the approximation. More review of meshless methods, we refer to \cite{nguyen2008meshless,chen2017meshfree}.

It is well known that the difficulty in solving PDEs arises from the differential operators, while the handling of other non-differential terms is relatively easy. How to describe the different differential operators is the central topic for different methods, including the meshless method, finite element method, finite difference method. The finite element method and meshless method use the shape function to interpolate the field value by the unknown nodal values, i.e. $u_h(\bx_a)=N_b(\bx_a) u_b$, where $N_b(\bx_a)$ is the shape function, $u_b$ is the nodal value. The differential operators correspond to the derivatives of the shape function, while the non-differential terms relate to the shape function. In certain cases, the construction of the shape function is complicated, let along the derivative calculation of the shape function. In this sense, it is desirable to jump into the derivatives directly while ignoring the shape functions.  {On the other hand, the method baseds on the local differential operator confronts inconveniences or difficulties when its definition does not exist for problems involving strong/weak discontinuity. For these problems, many meshless methods or finite element methods} (i.e. extended finite element method \cite{Belytschko19}) need special treatment to construct the shape function and to calculate the derivatives of the shape function. To circumvent the difficulties in methods based on local differential operators, many nonlocal theories have been proposed, among which include the nonlocal continuum field theories for different physical fields \cite{eringen2002nonlocal}, peridynamics \cite{Silling2000}, nonlocal integral form for plasticity and damage \cite{bavzant2002nonlocal} and nonlocal vector calculus \cite{gunzburger2010nonlocal}, to name a few. The nonlocal theory is based on the integral form with a finite intrinsic length scale, while the definition of a local differential operator is based on the intrinsic length scale approaching infinitesimal. Peridynamics proposed by Silling \cite{Silling2000} reformulates the elasticity theory into the integral form to account for the long range forces, which overcomes the difficulty to define the local derivatives for fractures. Comparing with the local theory, nonlocal theory not only has well-poseness in numerical aspect, but also approaches the real physical process better with an intrinsic length scale \cite{eringen2002nonlocal,Silling2000,bavzant2002nonlocal}.

{Mathematically, a nonlocal equation is a relation for which the information about the value of the function far from that point is required, in contrast with the differential equations describing relations between the values of an unknown function and its derivatives of different orders.} One common scenario for nonlocal equation is the equation involving integral operators, i.e.
\begin{align}
\frac{ \ud u}{ \ud t}(t,x)=\int_{\Omega} (u(t,y)-u(t,x))k(x,y) \ud y\label{eq:nonlocal1},
\end{align} 
for some kernel $k$, the integral operator is termed as nonlocal operator.

Another example is the nonlocal second-order scalar ``elliptic boundary-value'' problem,
\begin{align}
\mathcal L(u)(\bx):=2 \int_\Omega (u(\bx')-u(\bx)) k(\bx,\bx') \ud \bx'=b(\bx) \quad \mbox{ in }\Omega\label{eq:nonlocal2},
\end{align}
augmented with nonlocal ``Dirichlet'' or ``Neumann'' boundary condition, where $k(\bx,\bx')$ is the kernel function \cite{gunzburger2010nonlocal,du2013nonlocal}. The corresponded local form of Eq.\ref{eq:nonlocal2} is the second-order scalar elliptic boundary-value problem,
\begin{align}
-\nabla \cdot (\bD(\bx)\cdot \nabla u(\bx))=b(\bx) \quad \mbox{ in }\Omega\notag,
\end{align}
with Dirichlet or Neumann boundary conditions on the boundary $\partial \Omega$, where $\bD$ is a symmetric, positive definite, second-order tensor, $b$ a scalar-valued data function. {When the length scale  decreases to 0, the nonlocal form degenerates to the local form} \cite{du2013nonlocal}.

In the nonlocal equation, a point interacts with another point of finite distance, the intensity of interaction is related to the difference of field values of two points as indicated in Eq.\ref{eq:nonlocal1} and Eq.\ref{eq:nonlocal2}. The definition of the differential operator in PDEs resembles to the interaction between two points with finite distance. {We take the derivative of a scalar field for example, for vector or tensor field the Fr\'echet or G\^ateaux derivative can be applied. The derivative of scalar field }$u(x)$ {is defined as}
\[
u'(x)=\lim _{y\to x}{\frac {u(y)-u(x)}{y-x}}.
\]
The derivative is the limit of the difference of two points on their relative distance. Without seeking the limit, the nonlocal form based on the sum of weighted finite difference emerges naturally,
\[
\mathcal L(u)(x):=\int_{y \in \cS_{x}} {\frac {u(y)-u(x)}{y-x}} k(x,y) \ud y,
\]
where $k(x,y)$ is the weight function or kernel function, $\cS_x$ is the support. When the nonlocal length scale decreases to zero ($\cS_x\to 0$), $\mathcal L(u)(x)\to u'(x)$, which can be verified by Taylor series expansion in section \ref{ssec:nos}. Nonlocal operator provides a direct way to construct the differential operator, though the concept of ``locality'' is a special case of ``nonlocality''. Based on this basic observation, we construct several nonlocal differential operators to replace the local differential operators to solve PDEs. In the nonlocal operator method, the field value is defined on the node, therefore the use of the shape functions based on interpolation or approximation like FEM or meshless methods is no longer needed. The differential operator on field value is considered as the nonlocal interactions between the points in the support domain.

The purpose of the paper is to propose a {nonlocal operator method} for solving PDEs based on weighted residual method and variational principles. Though the nonlocal theory is more general than the local theory, we focus on solving the local problems with the nonlocal operator method. The nonlocal operator method constructs the nonlocal operator to represent the nonlocal interaction without shape function or its derivatives in traditional meshless or finite element method. The remainder of the paper is outlined as follows. In \S \ref{sec:support}, the concepts of support and dual-support are introduced. Based on the support, the general nonlocal operator and its variation in continuous form or discrete form are defined. In \S \ref{sec:hourglass}, we discuss the hourglass mode in the nonlocal operator and propose universal hourglass energy functional to remove the hourglass mode. The higher order nonlocal operators and higher order hourglass energy functional are generalized and obtained in \S \ref{sec:highhourglass}. Since the differential operator forms the basis of different energy functionals, we study the capabilities of the nonlocal operator based on variational principles in obtaining the strong forms or weak forms of different functionals in \S \ref{sec:vp}. Some numerical examples are presented to validate the method in \S \ref{sec:appli}. We conclude in \S \ref{sec:end}.
\section{Support, dual-support and nonlocal operators}\label{sec:support}
\begin{figure}[htp]
 \centering
 \subfigure[]{
 \label{fig:Coord}
 \includegraphics[width=.4\textwidth]{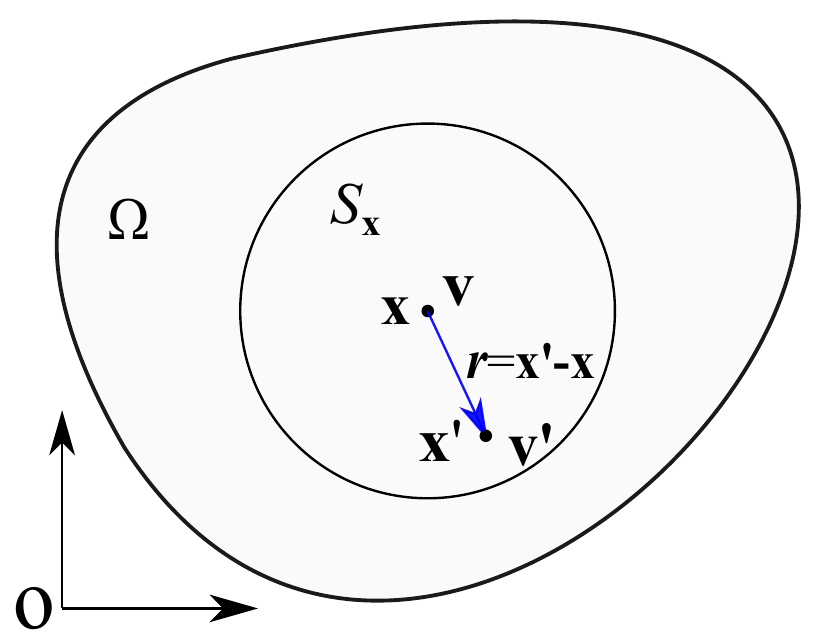}}
 \vspace{.1in}
 \subfigure[]{
 \label{fig:4support}
 \includegraphics[width=.35\textwidth]{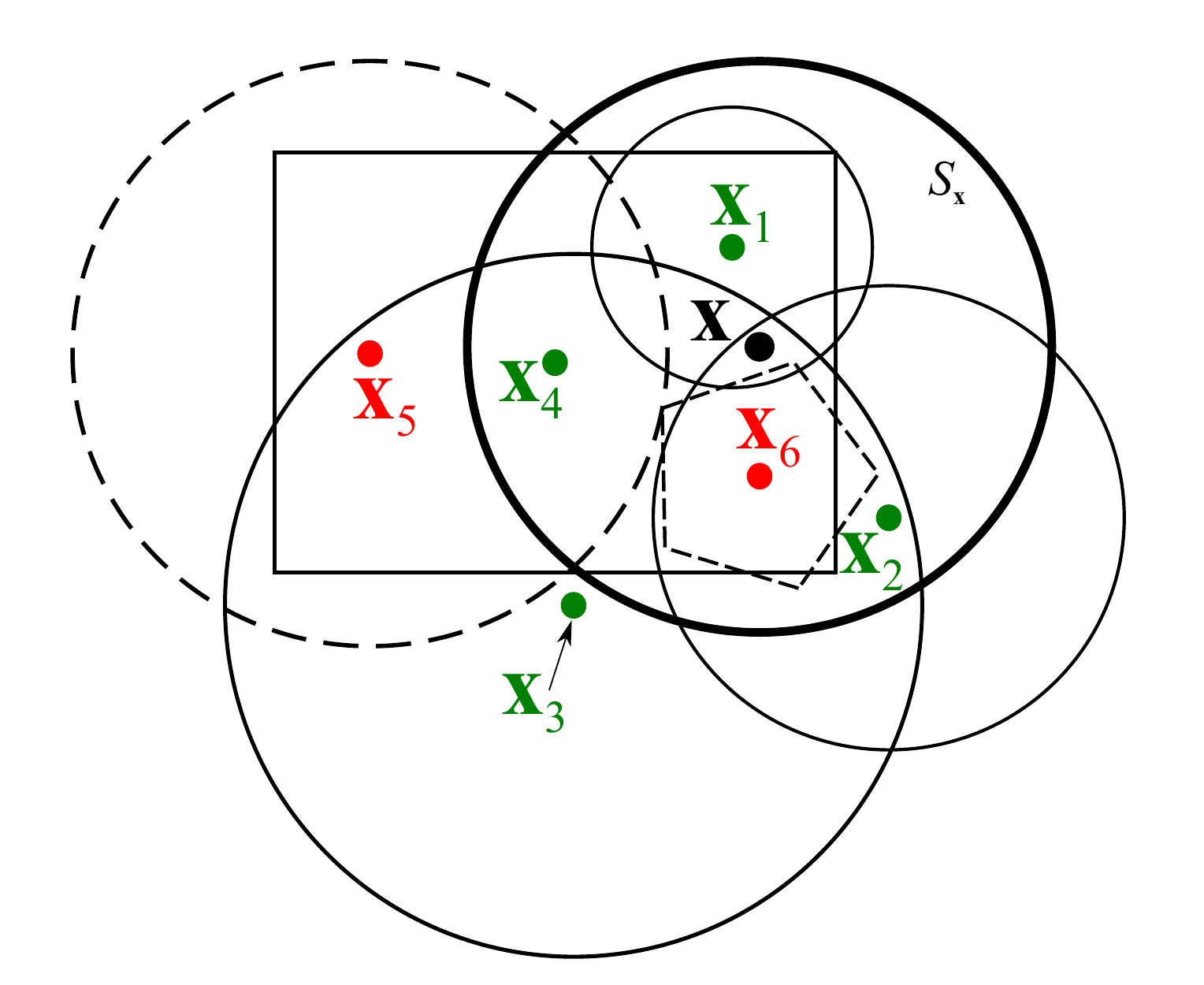}}\\
 \vspace{.3in}
\caption{(a) Domain and notations. (b) Schematic diagram for support and dual-support, all shapes above are support, $\cS_{\bx}=\{\bx_1,\bx_2,\bx_4,\bx_6\} $, $\cS_{\bx}'=\{\bx_1,\bx_2,\bx_3,\bx_4\}$.}
\end{figure} 
Consider a domain as shown in Fig.\ref{fig:Coord}, let $\bx$ be spatial coordinates in the domain $\mathbf \Omega$; $\br:=\bx'-\bx$ is a Euclidean vector ( or a spatial vector, or simply a vector) starts from $\bx$ to $\bx'$; $\bv:=\bv(\bx,t)$ and $\bv':=\bv(\bx',t)$ are the field values for $\bx$ and $\bx'$, respectively; $\bv_{\br}:=\bv'-\bv$ is the relative field vector for spatial vector $\br$.

\textbf{Support} $\cS_{\bx}$ of point $\bx$ is the domain where any spatial point $\bx'$ forms spatial vector $\br(=\bx'-\bx)$ from $\bx$ to $\bx'$. {Support} $\cS_{\bx}$ {specifies the range of nonlocal interaction happened with respect to point} $\bx$. {The main function of support is to define different nonlocal operators. In mathematics, a specific quantitative measure of the support can be described by a moment (shape tensor)}. A $n-$order moment of the support is defined as
\begin{align}
\bK_{\bx}^n:=\int_{\cS_{\bx}}w(\br) \underbrace{\br \otimes \br \otimes \cdots \otimes \br}_{\text{n terms}} \,\ud V_{\bx'},
\end{align}
where $w(\br)$ is the weight function.
Two special cases of the shape tensor for $\cS_{\bx}$ are the 0-order shape tensor (the weighted volume of the support) and the 2-order shape tensor 
\begin{align}
\bK_{\bx}:=\int_{\cS_{\bx}}w(\br) \br \otimes \br \,\ud V_{\bx'}.
\end{align}

\textbf{Dual-support} is defined as a union of the points whose supports include $\bx$, denoted by 
\begin{align}
\cS_{\bx}'=\{\bx'|\bx \in \cS_{\bx'}\} \label{eq:dualsupport}.
\end{align}
The point $\bx'$ forms dual-vector $\br'(=\bx-\bx'=-\br)$ in $\cS_{\bx}'$. On the other hand, $\br'$ is the spatial vector formed in $\cS_{\bx'}$. One example to illustrate the support and dual-support is shown in Fig.\ref{fig:4support}.

\subsection{Nonlocal operators in support}\label{ssec:nos}
The general operators in calculus include the gradient of scalar and vector field, the curl and divergence of vector field. {The definitions of some nonlocal operators can be found in reference} \cite{du2013nonlocal}. These operators have the corresponding nonlocal forms based on the Taylor series expansion. We use ${\tnabla}$ {to denote the nonlocal operator, while the local operators follow the conventional notations. \hl{The derivation adopts the conventions in classical calculus and linear algebra. The scalar, vector and two- or higher order tensor are denoted by small letter, small bold letter, capital bold letter, respectively.
The vector is based on column form. For the purpose of being concise, the following notations are used simultaneously, i.e.
$\mathbf a \cdot \mathbf b=\mathbf a^T \mathbf b$, 
$\mathbf a \otimes \mathbf b=\mathbf a \mathbf b^T$,
$\mathbf M \cdot \mathbf v=\mathbf M \mathbf v$,
$\mathbf v \cdot \mathbf M=\mathbf v^T \mathbf M$,
$\mathbf M_1\cdot \mathbf M_2=\mathbf M_1  \mathbf M_2$,
$\mathbf M_1 :\mathbf M_2=\mbox{tr}(\mathbf M_1 \mathbf M_2^T)$,
$\nabla =(\frac{\partial }{\partial x},\frac{\partial }{\partial y},\frac{\partial }{\partial z})^T$, and $\nabla \mathbf v=\nabla \otimes \mathbf v$, 
where superscript $T$ refers to the transpose operation.
The gradient of a scalar $v$ and 3-vector $\mathbf v=(v_1,v_2,v_3)^T$ are denoted by, respectively
\begin{align}
\nabla v=\begin{bmatrix}\frac{\partial v}{\partial x}\\\frac{\partial v}{\partial y}\\\frac{\partial v }{\partial z}\end{bmatrix},\quad
\nabla \mathbf v= \begin{bmatrix}\frac{\partial v_1 }{\partial x}&\frac{\partial v_1}{\partial y}&\frac{\partial v_1}{\partial z}\\\frac{\partial v_2 }{\partial x}&\frac{\partial v_2}{\partial y}&\frac{\partial v_2}{\partial z}\\\frac{\partial v_3 }{\partial x}&\frac{\partial v_3}{\partial y}&\frac{\partial v_3}{\partial z}\end{bmatrix}.\notag
\end{align}}

\textbf{Nonlocal gradient} of a vector field $\bv$ for point $\bx$ in support $\cS_{\bx}$ is defined as 
\begin{align}
\tnabla \otimes\bv_{\bx}:=\int_{\cS_{\bx}}w(\br) \bv_{\br} \otimes \br \,\ud V_{\bx'} \cdot \bK_{\bx}^{-1} \label{eq:FGdef},
\end{align}
where $\bv_{\br}=\bv_{\bx'}-\bv_{\bx}$, $\bK_{\bx}$ is the {2-order} shape tensor. {One example of the nonlocal gradient is the nonlocal deformation gradient in Peridynamics} \cite{Silling2007}.

In fact, the field value of nearby point $\bx'$ in $\cS_{\bx}$ is obtained by Taylor series expansion as
\begin{align}
\bv_{\bx'}=\bv_{\bx} +\nabla \otimes\bv_{\bx} \cdot \br+O(r^2)\label{eq:fTaylor},
\end{align}
where $O(r^2)$ represents order terms higher than one, and for linear field $O(r^2)=0$. Insert Eq.\ref{eq:fTaylor} into RHS of Eq.\ref{eq:FGdef} and integrate in support $\cS_\bx$, one verifies that the nonlocal operator converges to the local operator by the following derivation. 
\begin{align}
\tnabla\otimes \mathbf v_{\bx}=&\int_{\cS_{\bx}}w(\br) \bv_{\br} \otimes \br \,\ud V_{\bx'} \cdot \bK_{\bx}^{-1}\notag\\
=&\int_{\cS_{\bx}}w(\br) (\bv_\bx'-\bv_{\bx}) \otimes \br \,\ud V_{\bx'} \cdot \bK_{\bx}^{-1}\notag\\
=&\int_{\cS_{\bx}}w(\br) \nabla \mathbf v_{\bx} \cdot \br \otimes \br \,\ud V_{\bx'} \cdot \bK_{\bx}^{-1}\notag\\
=&\nabla \otimes\mathbf v_{\bx} \cdot\int_{\cS_{\bx}}w(\br) \br \otimes \br \,\ud V_{\bx'} \cdot \bK_{\bx}^{-1}\notag\\
=&\nabla \otimes\mathbf v_{\bx} \cdot\bK_{\bx} \cdot \bK_{\bx}^{-1}\notag\\
=&\nabla \otimes\mathbf v_{\bx}.\notag
\end{align}
When $\bx'$ is close enough to $\bx$ or when support $\cS_\bx$ is small enough, the nonlocal operator can be considered as the linearization of the field. The nonlocal operator converges to the local operator in the continuous limit. On the other hand, the nonlocal operator defined by integral form, still holds in the case where strong discontinuity exists. The local operator can be viewed as a special case of the nonlocal operator.

Similarly, \textbf{nonlocal gradient} of a scalar field $v$ for point $\bx$ in support $\cS_{\bx}$ is defined as 
\begin{align}
\tnabla {v}_{\bx}:=\int_{\cS_{\bx}}w(\br) v_{\br} \br \,  \ud V_{\bx'} \cdot \bK_{\bx}^{-1} \label{eq:sFGdef},
\end{align}
where $v_{\br}=v_{\bx'}-v_{\bx}$.

Let $[\square]_\times$ denote the map of a $3\times 3$ antisymmetric matrix into 3-vector,
\begin{align}
\begin{bmatrix}0&a_{3}&-a_{2}\\ -a_{3}&0&a_{1}\\a_{2}&-a_{1}&0\end{bmatrix}_{\times }\mapsto(a_1,a_2,a_3)^T
\end{align}
It is easy to verify that for any vector $\mathbf u$
\begin{align}
\nabla \times \mathbf u=[\nabla \otimes \mathbf u-(\nabla \otimes \mathbf u)^T]_\times \ ,
\label{eq:curlnabla} 
\end{align}
Now if we follow the nonlocal gradient operator for $\mathbf v$ according to  Eq.\ref{eq:FGdef} , Eq. \ref{eq:curlnabla} will become
\begin{align}
&\tnabla \times \bv=[\tnabla \otimes \bv-(\tnabla \otimes \bv)^T]_\times\notag\\
&=\left [ \int_{\cS_{\bx}}w(\br) \bv_{\br} \otimes (\bK_{\bx}^{-1}\cdot\br) \, \mathrm{d} V_{\bx'}  -\int_{\cS_{\bx}}w(\br)  (\bK_{\bx}^{-1}\cdot\br) \otimes\bv_{\br} \, \mathrm{d} V_{\bx'} \right ]_\times \notag\\
&=\left [\int_{\cS_{\bx}}w(\br) \left (\bv_{\br} \otimes (\bK_{\bx}^{-1}\cdot\br) -  (\bK_{\bx}^{-1}\cdot\br) \otimes\bv_{\br} \right) \, \mathrm{d} V_{\bx'} \right ]_\times\notag \\
&=\int_{\cS_{\bx}}w(\br) \big[\bv_{\br} \otimes (\bK_{\bx}^{-1}\cdot\br) -  (\bK_{\bx}^{-1}\cdot\br) \otimes\bv_{\br}\big]_\times \mathrm{d} V_{\bx'}\notag  \\
&=\int_{\cS_{\bx}}w(\br) (\mathbf K_{\bx}^{-1} \cdot\br)\times \bv_{\br} \mathrm{d} V_{\bx'}\notag
\end{align}
In the last step, $\mathbf b\times \mathbf a=\big[\mathbf a \otimes \mathbf b-\mathbf b\otimes \mathbf a\big]_\times $ is used. 

Hence, \textbf{nonlocal curl} of a vector field $\bv$ for point $\bx$ in support $\cS_{\bx}$ is defined as 
\begin{align}
\tnabla \times \bv_\bx:=\int_{\cS_{\bx}}w(\br) (\mathbf K_{\bx}^{-1} \cdot\br)\times \bv_{\br} \,\ud V_{\bx'}\label{eq:Fdef}.
\end{align}

By analogy with $\nabla \cdot \bv=\mbox{tr}(\nabla \otimes \bv)$, where $\mbox{tr}(\square)$ denotes the trace of the matrix, the nonlocal divergence of a vector field $\bv$ for point $\bx$ in support $\cS_{\bx}$ is derived as 
\begin{align}
&\tnabla \cdot \bv_{\bx}=\mbox{tr}(\tnabla \otimes \bv)\notag\\
&=\mbox{tr}(\int_{\cS_{\bx}}w(\br) \bv_{\br} \otimes (\bK_{\bx}^{-1}\cdot\br) \,\ud V_{\bx'})\notag\\
&=\int_{\cS_{\bx}}w(\br) \mbox{tr}\big(\bv_{\br} \otimes (\bK_{\bx}^{-1}\cdot\br)\big) \,\ud V_{\bx'}\notag\\
&=\int_{\cS_{\bx}}w(\br) \bv_{\br} \cdot (\bK_{\bx}^{-1}\cdot\br) \,\ud V_{\bx'}\notag
\end{align}
In the third step, $\mbox{tr}(\mathbf a\otimes \mathbf b)=\mathbf a\cdot \mathbf b$ is used. 

Hence, \textbf{nonlocal divergence} is defined as
\begin{align}
\tnabla \cdot \bv_{\bx}:=\int_{\cS_{\bx}}w(\br) \bv_{\br} \cdot (\bK_{\bx}^{-1}\cdot\br) \,\ud V_{\bx'}\label{eq:Fddef}.
\end{align}
In the continuous limit, based on the nonlocal gradient, the nonlocal curl and nonlocal divergence converge to the conventional curl and divergence operator, respectively.
\subsection{Variation of the nonlocal operator}
We present the variation of the general nonlocal operator in continuous form and discrete form. The discrete form is beneficial for the numerical implementation. Different nonlocal operators can be used to replace the local differential operators in PDEs, {especially in the framework of weighted residual method and variational principles}. We use the $\delta$ to denote the variation. 
\subsubsection{Notations for variation}
The nonlocal operators defined above are in vector or tensor form. The variation of the nonlocal operators leads to a higher-order tensor form, which is not convenient for implementation. We need to express the higher order tensor into to vector or matrix form. 
Before we derive the variation of nonlocal operator, some notation to denote the variation and how the variations are related to the first- and second-order derivatives are to be discussed. Assuming a functional $\fF(u,v)$, where $u,v$ are unknown functions in unknown vector $[u,v]$, the first and second variation can be expressed as
\begin{align}
\delta \fF(u,v)&=\partial_u \fF \delta u+\partial_v \fF\delta v=[\partial_u \fF,\partial_v \fF]\begin{bmatrix}\delta u\\\delta v\end{bmatrix}\notag\\
\delta^2 \fF(u,v)&=\partial_{uu} \fF \delta u\delta u+\partial_{uv} \fF \delta u\delta v+\partial_{vu} \fF\delta v \delta u+\partial_{vv} \fF\delta v \delta v\notag\\
&=\begin{bmatrix}\partial_{uu} F & \partial_{uv} \fF\\ \partial_{vu} \fF& \partial_{vv} \fF\end{bmatrix}:\begin{bmatrix}\delta u \delta u& \delta u\delta v\\\delta v\delta u&\delta v\delta v\end{bmatrix}\notag\\
\mbox{where }&\begin{bmatrix}\delta u \delta u& \delta u\delta v\\\delta v\delta u&\delta v\delta v\end{bmatrix}=\begin{bmatrix}\delta u\\\delta v\end{bmatrix}\otimes \begin{bmatrix}\delta u&\delta v\end{bmatrix}\notag
\end{align}
It can be seen that, the second variation $\delta^2\fF(u,v)$ is the double inner product of the Hessian matrix and the tensor formed by the variation of the unknowns, while the first variation $\delta\fF(u,v)$ is inner product of the gradient vector and the variation of the unknowns. The gradient vector and Hessian matrix represent the residual vector and tangent stiffness matrix of the functional, respectively, with unknown functions $u,v$ being the independent variables,
\begin{align}
&\mathbf R=\nabla_{[u,v]}\fF(u,v)=[\partial_u\fF,\partial_v\fF]\notag\\
&\mathbf K=\nabla^2_{[u,v]}\fF(u,v)=\begin{bmatrix}\partial_{uu}\fF & \partial_{uv}\fF\\ \partial_{vu}\fF& \partial_{vv}\fF\end{bmatrix}.\notag
\end{align}
The inner product or double inner product indicates that location of an element in the residual or the tangent stiffness matrix corresponds to the location of the unknowns with variation. 

In this paper, we use a special variation $\tdelta$, whose function is illustrated by the following examples. The special variations of functional $F(u,v)$ are given as
\begin{align}
\tdelta\fF(u,v)&=\partial_u\fF \tdelta u+\partial_v\fF\tdelta v=[\partial_u\fF,\partial_v\fF]\notag\\
\tdelta^2\fF(u,v)&=\partial_{uu}\fF \tdelta u\tdelta u+\partial_{uv}\fF \tdelta u\tdelta v+\partial_{vu}\fF\tdelta v \tdelta u+\partial_{vv}\fF\tdelta v \tdelta v\notag\\
&=\begin{bmatrix}\partial_{uu}\fF & \partial_{uv}\fF\\ \partial_{vu}\fF& \partial_{vv}\fF\end{bmatrix}\notag
\end{align}
where $\tdelta u$ denotes the index of $\partial_u\fF$ in residual vector by the index of $u$ in the unknown vector. For example, the term $\partial_v\fF\tdelta v$ represents $\partial_v\fF$ be in the second location of the residual vector since $v$ is in the second position of $[u,v]$. The term $\partial_{uv}\fF \tdelta u\tdelta v$ denotes that the location of $\partial_{uv}\fF$ is (1,2), while the term $\partial_{vu}\fF \tdelta v\tdelta u$ denotes that the location of $\partial_{vu}\fF$ is (2,1).

Obviously,
\begin{align}
&\delta\fF(u,v)=\tdelta\fF(u,v)\begin{bmatrix}\delta u\\\delta v\end{bmatrix}, \delta^2\fF(u,v)=\tdelta^2\fF(u,v):\begin{bmatrix}\delta u \delta u& \delta u\delta v\\\delta v\delta u&\delta v\delta v\end{bmatrix}\notag\\
&\mathbf K=\tdelta^2\fF(u,v),\mathbf R=\tdelta\fF(u,v)\notag
\end{align}
The special first-order and second-order variation of a functional lead to the residual and tangent stiffness matrix directly. The traditional variation can be recovered by the inner product of the special variation and the variation of the unknown vector. 
\subsubsection{Nonlocal divergence operator}
The variation of $\tnabla \cdot \bF_{\bx}$ is given by
\begin{align}
\tnabla \cdot \delta \mathbf F_{\bx}=\int_{\cS_{\bx}} w(\br) (\bK_{\bx}^{-1}\cdot\br) \cdot (\delta \mathbf F_{\bx'}-\delta \mathbf F_{\bx}) \,\ud V_{\bx'},\label{eq:dvaF}
\end{align}
The number of dimensions of $\tnabla \cdot \delta \mathbf F_{\bx}$ is infinite, and discretization is required.

After discretization of the domain by particles, the whole domain is represented by
\begin{align}
\Omega=\sum_{i=1}^{Nnode} \Delta V_i
\end{align}
where $i$ is the global index of volume $\Delta V_i$, $Nnode$ is the number of particles in $\Omega$.

Particles in $\cS_{i}$ are represented by 
\begin{align}
N_{i}=\{i,j_1,..,j_k,..,j_{n_{i}}\}\label{eq:Ni}
\end{align}
where $j_1,..,j_k,..,j_{n_{i}}$ are the global indices of neighbors of particle $i$. 
The discrete form of $\tnabla \cdot \delta \mathbf F_{i}$ can be written as
\begin{align}
\tnabla \cdot \delta \mathbf F_{i}\simeq\sum_{j_k\in \cS_{i}} w(\br) \Delta V_{j_k} (\mathbf K_{i}^{-1} \br) \cdot (\delta \mathbf F_{j_k}-\delta \mathbf F_{i})=\tnabla \cdot \tdelta \mathbf F_{i}\cdot \delta \bF_{N_i},\label{eq:dvaF2}
\end{align}
where $\simeq$ denotes discretization, $\delta \bF_{N_i}$ is all the variations of the unknowns in support $\cS_i$,
\begin{align}
\delta \bF_{N_i}&=(\delta \bF_i,\delta \bF_{j_1},..,\delta \bF_{j_k},..,\delta \bF_{j_{n_i}}),\label{eq:deltaFni}
\end{align}
$\tnabla \cdot \tdelta \mathbf F_{i}$ is the coefficient vector with a length of $3 (n_i+1)$ in 3D case,
\begin{align}
\tnabla \cdot \tdelta \mathbf F_{i}=\sum_{j_k\in \cS_{i}} w(\br) \Delta V_{j_k} (\mathbf K_{i}^{-1} \br) \cdot (\tdelta \mathbf F_{j_k}-\tdelta \mathbf F_{i}).
\end{align}
Based on the indices of $\tdelta \mathbf F_{j_k}$ in $\delta \bF_{N_i}$, $\tnabla \cdot \tdelta \mathbf F_{i}$ can be obtained by
\begin{align}
\tnabla \cdot \tdelta \mathbf F_{i}{[3 k,3k+1,3k+2]}&=w(\br) \Delta V_{j_k} \bK_{i}^{-1} \br, \,\tnabla \cdot \tdelta \mathbf F_{i}{[0,1,2]}=-\sum_{k=1}^{n_i}w(\br) \Delta V_{j_k} \bK_{i}^{-1} \br,\label{eq:deltaDivF}
\end{align}
where $k$ is the index of particle $j_k$ in $N_{i}$. The process to obtain $\tnabla \cdot \tdelta \mathbf F_{i}$ on the nodal level is called nodal assembly. In the following section, we mainly discuss the special variation of the nonlocal operator and functional, while the actual variation can be recovered with ease.

\subsubsection{Nonlocal curl operator}
The variation of $\tnabla \times \bF_{i}$ in discrete form reads
\begin{align}
\tnabla\times \tdelta \mathbf F_{i}\simeq\sum_{j_k\in \cS_{i}} w(\br) \Delta V_{j_k} \mathbf K_{i}^{-1}\cdot \br \times (\tdelta \mathbf F_{j_k}-\tdelta \mathbf F_{i}),\label{eq:cvaF}
\end{align}
where $\Delta V_{j_k}$ is the volume for particle $j_k$. For the 3D case, $\tnabla\times \tdelta \mathbf F_{i}$ is a $3 \times 3 (n_{i}+1)$ matrix, where $n_{i}$ is the number of neighbors in $\cS_{i}$, $N_i$ is given by Eq.\ref{eq:Ni} . For each particle $j_k$ in $N_{i}$ calculating $R_{j_k}=w(\br) \Delta V_{j_k} \bK_{i}^{-1} \br$, we obtain
\begin{align}
&\tnabla\times \tdelta \mathbf F_{i}{[1,3 k]}=R_{j_k}[2], \tnabla\times \tdelta \mathbf F_{i}{[2,3 k]}=-R_{j_k}[1] ,\tnabla\times \tdelta \mathbf F_{i}{[0,3 k+1]}=-R_{j_k}[2]\notag\\
&\tnabla\times \tdelta \mathbf F_{i}{[2,3 k+1]}=R_{j_k}[0],\tnabla\times \tdelta \mathbf F_{i}{[0,3 k+2]}=R_{j_k}[1],\tnabla\times \tdelta \mathbf F_{i}{[1,3 k+2]}=-R_{j_k}[0]\notag\\
&\tnabla\times \tdelta \mathbf F_{i}{[1,0]}=-\sum_{k=1}^{n_i} R_{j_k}[2], \tnabla\times \tdelta \mathbf F_{i}{[2,0]}=\sum_{k=1}^{n_i} R_{j_k}[1],\tnabla\times \tdelta \mathbf F_{i}{[0,1]}=\sum_{k=1}^{n_i} R_{j_k}[2],\notag\\
&\tnabla\times \tdelta \mathbf F_{i}{[2,1]}=-\sum_{k=1}^{n_i} R_{j_k}[0], \tnabla\times \tdelta \mathbf F_{i}{[0,2]}=-\sum_{k=1}^{n_i} R_{j_k}[1], \tnabla\times \tdelta \mathbf F_{i}{[1,2]}=\sum_{k=1}^{n_i} R_{j_k}[0],\label{eq:deltaFx}
\end{align}
where $k$ is the index of particle $j_k$ in $N_{i}$.  The minus sign denotes the reaction from the dual-support, which guarantees the regularity of the stiffness matrix in the absence of external constraints. The nodal assembly for the variation of the vector cross product can be obtained by
\begin{align}
\mathbf F^\times=\{R_0,R_1,R_2\}\times\{F_0,F_1,F_2\}=\{{F_2} {R_1}-{F_1} {R_2},{F_0} {R_2}-{F_2} {R_0},{F_1} {R_0}-{F_0} {R_1}\}
\end{align}
while the gradient of $\mathbf F^\times$ on $\{F_0,F_1,F_2\}$ is given by
\begin{align}\begin{bmatrix}
\frac{\partial F^\times_0}{\partial F_0} & \frac{\partial F^\times_0}{\partial F_1} & \frac{\partial F^\times_0}{\partial F_2} \\
\frac{\partial F^\times_1}{\partial F_0} & \frac{\partial F^\times_1}{\partial F_1} & \frac{\partial F^\times_1}{\partial F_2} \\
\frac{\partial F^\times_2}{\partial F_0} & \frac{\partial F^\times_2}{\partial F_1} & \frac{\partial F^\times_2}{\partial F_2} \\
\end{bmatrix}=
\begin{bmatrix}
 0 & -{R_2} & {R_1} \\
 {R_2} & 0 & -{R_0} \\
 -{R_1} & {R_0} & 0 \\
\end{bmatrix}.\label{eq:curlIndex}
\end{align}
The indices of $R$ correspond to their locations in $\mathbf F^{\times}$.\\
\subsubsection{Nonlocal gradient operator for vector field}
Similarly, the variation of $\tnabla \otimes\bF_{i}$ in the discrete form reads
\begin{align}
\tnabla\otimes\tdelta \bF_i\simeq\sum_{j_k\in \cS_{i}} w(\br) (\tdelta \bF_{j_k}-\tdelta \bF_{i})\otimes (\bK_{i}^{-1} \br) \Delta V_{j_k}
\end{align}
where $\Delta V_{j_k}$ is the volume for particle $j_k$. In 3D, $\tnabla\tdelta \bF_i$ is a $9 \times 3 (n_{i}+1)$ matrix, where $n_{i}$ is the number of neighbors in $\cS_{i}$, $N_i$ is given by Eq.\ref{eq:Ni}. For each particle in the neighbor list with $R_{j_k}=w(\br) \Delta V_{j_k} \bK_{i}^{-1} \br$, the terms in $R_{j_k}$ can be added to the $\tnabla\otimes\tdelta \bF_i$ as 
\begin{align}
&\tnabla\otimes\tdelta \bF_i{[0,3 k]}=R_{j_k}[0],  
\tnabla\otimes\tdelta \bF_i{[3,3 k]}=R_{j_k}[1],  
\tnabla\otimes\tdelta \bF_i{[6,3 k]}=R_{j_k}[2],\notag\\  
&\tnabla\otimes\tdelta \bF_i{[1,3 k+1]}=R_{j_k}[0],
\tnabla\otimes\tdelta \bF_i{[4,3 k+1]}=R_{j_k}[1],
\tnabla\otimes\tdelta \bF_i{[7,3 k+1]}=R_{j_k}[2],\notag\\
&\tnabla\otimes\tdelta \bF_i{[2,3 k+2]}=R_{j_k}[0],
\tnabla\otimes\tdelta \bF_i{[5,3 k+2]}=R_{j_k}[1],
\tnabla\otimes\tdelta \bF_i{[8,3 k+2]}=R_{j_k}[2],\notag\\
&\tnabla\otimes\tdelta \bF_i{[0,0]}=-\sum_{k=1}^{n_i}R_{j_k}[0],
\tnabla\otimes\tdelta \bF_i{[3,0]}=-\sum_{k=1}^{n_i}R_{j_k}[1],
\tnabla\otimes\tdelta \bF_i{[6,0]}=-\sum_{k=1}^{n_i}R_{j_k}[2],\notag\\
&\tnabla\otimes\tdelta \bF_i{[1,1]}=-\sum_{k=1}^{n_i}R_{j_k}[0],
\tnabla\otimes\tdelta \bF_i{[4,1]}=-\sum_{k=1}^{n_i}R_{j_k}[1],
\tnabla\otimes\tdelta \bF_i{[7,1]}=-\sum_{k=1}^{n_i}R_{j_k}[2],\notag\\
&\tnabla\otimes\tdelta \bF_i{[2,2]}=-\sum_{k=1}^{n_i}R_{j_k}[0],
\tnabla\otimes\tdelta \bF_i{[5,2]}=-\sum_{k=1}^{n_i}R_{j_k}[1],
\tnabla\otimes\tdelta \bF_i{[8,2]}=-\sum_{k=1}^{n_i}R_{j_k}[2],\label{eq:deltaDF}
\end{align}
where $k$ is the index of particle $j_k$ in $N_{i}$. The sub-index of $\tnabla \tdelta \bF_{\bx}$ can be obtained by the way similar to Eq.(\ref{eq:curlIndex}).
\subsubsection{Nonlocal gradient operator for scalar field}
The variation of $\nabla v_{i}$ reads
\begin{align}
\tnabla \tdelta v_{i}\simeq\sum_{ \cS_{i}} w(\br) \Delta V_{j_k} \mathbf K_{i}^{-1} \br (\tdelta v_{j_k}-\tdelta v_{i})\label{eq:vaFs},
\end{align}
where $\Delta V_{\bx'}$ is the volume for particle $\bx'$. For 3D case, the dimensions of $(\tnabla \tdelta v_{i})$ are $ 3 \times (n_{i}+1)$, where $n_{i}$ is the number of neighbors in $\cS_{i}$, $N_i$ is given by Eq.\ref{eq:Ni}. For each particle in the neighbor list with $R_{j_k}=w(\br) \Delta V_{j_k} \bK_{i}^{-1} \br$, the terms in $R_{j_k}$ can be added to the $\tnabla\tdelta \bF_i$ as 
\begin{align}
&\tnabla \tdelta v_{i}{[0,k]}=R_{j_k}[0],
\tnabla \tdelta v_{i}{[1,k]}=R_{j_k}[1],
\tnabla \tdelta v_{i}{[2,k]}=R_{j_k}[2],\notag\\
 &\tnabla \tdelta v_{\bx}{[0,0]}=-\sum_{k=1}^{n_i}R_{j_k}[0],
\tnabla \tdelta v_{\bx}{[1,0]}=-\sum_{k=1}^{n_i}R_{j_k}[1],
\tnabla \tdelta v_{\bx}{[2,0]}=-\sum_{k=1}^{n_i}R_{j_k}[2],\label{eq:gradVari2}
\end{align}
where $k$ is the index of particle $j_k$ in $N_{i}$. 

It can be seen that the basic element in the nodal assembly is $w(\br) \Delta V_{j_k} \bK_{i}^{-1} \br$, which is quite similar to the derivative of shape function in meshless or finite element method.
\section{Hourglass energy functional}\label{sec:hourglass}
\begin{figure}[htp]
	\centering
		\includegraphics[width=9cm]{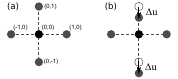}
	\caption{Hourglass mode demonstration. (a) initial configuration. (b) up and down particles with same rigid translation $\Delta u$. The deformation gradients defined by (a) and (b) are the same.}
	\label{fig:hgDemo}
\end{figure}

The same nonlocal operator can be defined by several configurations, for example, the initial configuration Fig.\ref{fig:hgDemo}(a) with a rigid translation $\Delta u$ for the up and down particles turns into Fig.\ref{fig:hgDemo}(b). It is easy to verify that the nonlocal gradient of $\bu$ from Eq.\ref{eq:FGdef} is zero, the same as that in the initial configuration. It can be seen that for the same field gradient, the configuration is not unique, where the extra deformation not accounted by the gradient is called the hourglass mode, which has zero energy contribution for the energy functional. The hourglass mode in the nonlocal operator is due to the deformation vectors counteracting with each other during the summation. 

In order to remove the hourglass mode (zero-energy mode), we propose a penalty energy functional to achieve the linear field of the vector field. The penalty energy functional is defined as the weighted difference square between current value of a point and the value predicted by the field gradient. In fact, the vector field in the neighborhood of a particle is required to be linear. Therefore, it has to be exactly described by the gradient of the vector field, and the hourglass modes are identified as that part of the vector field, which is not described by the vector field gradient. In practice, the difference of current deformed vector $\bv_{\br}$ and predicted vector by field gradient ($\mathbf{F}_{\bx}(=\nabla \bv)$ in Eq.\ref{eq:FGdef}) is $(\bF\br-\bv_{\br})$. We formulate the hourglass energy based on the difference in support as follows. Let $\alpha=\frac{\mu}{m_{\bK}}$ be a coefficient for the hourglass energy, where ${m_{\bK}}=\tr{\bK}$, $\mu$ is the penalty coefficient, the functional for zero-energy mode is
\begin{align}
\fF^{hg}&=\alpha \int_{\cS} w(\br) (\bF\br-\bv_{\br})^T (\bF\br-\bv_{\br}) \,\ud V\notag\\
&=\alpha \int_{\cS} w(\br)\Big(\br^T \bF^T \bF\br+\bv_{\br}^T \bv_{\br}-2 \bv_{\br}^T \bF\br\Big) \,\ud V\notag\\
&=\alpha \int_{\cS} w(\br)\Big(\bF^T \bF: \br \otimes \br+\bv_{\br}^T \bv_{\br}-2 \bF:\bv_{\br}\otimes\br\Big) \,\ud V\notag\\
&=\alpha \bF^T \bF: \int_{\cS} w(\br)\br \otimes \br \,\ud V+\alpha \int_{\cS} w(\br) \bv_{\br}^T \bv_{\br} \,\ud V-2 \alpha \bF:\int_{\cS} w(\br)\bv_{\br}\otimes\br \,\ud V\notag\\
&=\alpha \bF^T \bF: \bK+\alpha \int_{\cS} w(\br) \bv_{\br}\cdot \bv_{\br} \,\ud V-2 \alpha \bF: (\bF \bK)\notag\\
&=\frac{\mu}{m_{\bK}} \Big(\int_{\cS} w(\br) \bv_{\br}\cdot \bv_{\br} \,\ud V- \bF: \bF\bK\Big).
\end{align}
The above definition of hourglass energy is similar to the variance in probability theory and statistics. In above derivation, we used the relations:
$ \bF^T \bF: \bK=\bF: (\bF \bK), \mathbf a^T \mathbf M \mathbf b=\mathbf M: \mathbf a\otimes \mathbf b, \mathbf A:\mathbf B=\tr{\mathbf A \mathbf B^T}$, where capital letter denotes matrix and the small letter is column vector. The purpose of $m_{\bK}$ is to make the energy functional independent with the support since shape tensor $\bK$ is involved in $\bF^T \bF: \bK$. It should be noted that the zero-energy functional is valid in any dimensions and there is no limitation on the shape of the support.

The variation of $\tdelta(\bF: \bF\bK)$ can be rewritten as
\begin{align}
\tdelta(\bF:\bF\bK)&=\tdelta(\bF\bK:\bF)=2 \bF\bK:\tdelta \bF\notag\\
&=2 \bF\bK:\int_{\cS} w(\br) \tdelta \bv_{\br}\otimes (\bK^{-1} \br)\,\ud V\notag\\
&=2 \int_{\cS} w(\br) \tdelta \bv_{\br}^T \bF\bK (\bK^{-1} \br)\,\ud V\notag\\
&=2 \int_{\cS} w(\br) \tdelta \bv_{\br}^T (\bF \br)\,\ud V\notag\\
&=2 \int_{\cS} w(\br) (\bF \br)\cdot \tdelta\bv_{\br}\,\ud V.
\end{align}
Then the variation of $\fF^{hg}$ is
\begin{align}
\mathbf R^{hg}&=\tdelta\fF^{hg}\notag\\
&=\frac{\mu}{m_{\bK}} \Big(\int_{\cS} w(\br)\tdelta( \bv_{\br}\cdot \bv_{\br}) \,\ud V-\tdelta( \bF: \bF\bK)\Big)\notag\\
&=\frac{\mu}{m_{\bK}} \Big(\int_{\cS}2 w(\br) \bv_{\br}\cdot \tdelta\bv_{\br} \,\ud V-2 \int_{\cS} w(\br) (\bF \br)\cdot \tdelta \bv_{\br}\,\ud V\Big)\notag\\
&=\frac{2\mu}{m_{\bK}} \int_{\cS} w(\br) (\bv_{\br}-\bF \br)\cdot (\tdelta\bv'-\tdelta\bv) \,\ud V\label{eq:Phg}.
\end{align}
$\mathbf R^{hg}$ is the residual for hourglass energy. For a consistent field, the residual for hourglass mode is zero.

The hourglass control for individual spatial vector $\br$ can be written as
\begin{align}
\mathbf T^{hg}_{\br}=\fracp{\fF^{hg}_{\br}}{\bv}=\frac{2\mu}{m_{\bK}} w(\br)\big(\bv_{\br}-\bF_{\bx}\br\big).\label{eq:Phg20}
\end{align}
Eq.\ref{eq:Phg20} gives the explicit formula for the hourglass force. The term on $\tdelta \bv$ is the hourglass term from its support, while the terms on $\tdelta \bv'$ are the hourglass terms for the dual support $\cS_{\bx'}'$ of point $\bx'$. The second variation of $\fF^{hg}$ at a point is the tangent stiffness matrix,
\begin{align}
\mathbf K^{hg}=\tdelta^2\fF^{hg}=\frac{2\mu}{m_{\bK}} \Big(\int_{\cS} w(\br) (\tdelta \bv'-\tdelta \bv)^T (\tdelta \bv'-\tdelta \bv) \,\ud V-\tdelta \bF \bK: \tdelta \bF\Big).
\end{align}
The second variation of the zero-energy functional is the stiffness matrix, which is constant and can solve the rank deficiency in {nodal integration method}.

The above equations indicate that when the unknown field is consistent with the field gradient, the hourglass energy residual is zero. In this sense, hourglass energy functional contributes to the linear completeness.
\section{Higher order nonlocal operators and hourglass energy functional}\label{sec:highhourglass}
\subsection{Higher order nonlocal operators}
For certain problem like the plate/shell theory, the Hessian matrix of scalar field is required. In this case, the 2-order nonlocal Hessian operator and its variation are necessary. In this section, we extend the linear nonlocal operator to higher order ones.

For a scalar field $u$, the Taylor series expansion can be written as
\begin{align}
u_{\br}=\nabla u\cdot \br+\frac{1}{2!}\nabla^2 u:\br^2+\frac{1}{3!}\nabla^3 u\,\vdots\,\br^3+\cdots+\frac{1}{n!}\nabla^n u\cdot^{(n)}\br^n,
\end{align}
where $\cdot^{(2)}=:\,, \cdot^{(3)}=\vdots$ and $\cdot^{(n)}$ is the generalization of higher order inner product, 
\[
\br^n:=\underbrace{\br\otimes \br\otimes\cdots\otimes \br}_{\text{n terms}}.
\]

The $n$-order shape tensor is rewritten as 
\begin{align}
\bK_n=\int_{\cS} w(\br) \br^n \,\ud V.
\end{align}
For simplicity, we consider the Hessian nonlocal operator
\begin{align}
\frac{1}{2!}\nabla^2 u:\br^2= u_{\br}-\nabla u\cdot \br\label{eq:grad2}
\end{align}
The sum of weighted tensor of Eq.\ref{eq:grad2} multiplying $\br^2$ in support is 
\begin{align}
\frac{1}{2!}\nabla^2 u:\int_{\cS} w(\br)\br^4 \,\ud V=\int_{\cS} w(\br) (u_{\br}\br^2-\nabla u\cdot \br^3)d V.
\end{align}
The weighted sum can be simplified as follows.
\begin{align}
\frac{1}{2!}\nabla^2 u:\bK_4&=\int_{\cS} w(\br) (u_{\br}\br^2-\nabla u\cdot \br^3) \ud V\notag\\
&=\sum_{\cS} w(\br) \Delta V' (u_{\br}\br^2-\nabla u\cdot \br^3)\notag\\
&=\sum_{\cS} w(\br) \Delta V'u_{\br}\br^2-\nabla u\cdot \sum_{\cS} w(\br) \Delta V' \br^3\notag\\
&=\sum_{\cS} w(\br) \Delta V'u_{\br}\br^2-\sum_{\cS} w(\br) \Delta V' u_{\br} \br \cdot \bK_2^{-1}\cdot \bK_3\notag\\
&=\sum_{\cS} w(\br) \Delta V'u_{\br}\br^2-\sum_{\cS} w(\br) \Delta V' u_{\br} \bK_3 \bK_2^{-1} \br\notag\\
&=\sum_{\cS} w(\br) \Delta V'u_{\br}(\br^2- \bK_3 \bK_2^{-1} \br)\notag\\
&=\int_{\cS} w(\br) u_{\br}(\br^2- \bK_3 \bK_2^{-1} \br) \ud V.
\end{align}
Hence, the nonlocal Hessian operator and its variation can be written as 
\begin{align}
\tnabla^2 u &=2!\int_{\cS} w(\br) u_{\br}(\br^2- \bK_3 \bK_2^{-1} \br) \ud V : \bK_4^{-1}\\
\tnabla^2 \tdelta u &=2!\int_{\cS} w(\br) (\tdelta u'-\tdelta u)(\br^2- \bK_3 \bK_2^{-1} \br) \ud V : \bK_4^{-1}.
\end{align}
Examples to calculate the nonlocal Hessian operator in 1D and 2D are given in \ref{app:h}.
\subsection{Higher order hourglass energy functional}
Consider a $n$-order hourglass energy functional defined by
\begin{align}
\fF_n^{hg}=\alpha\int_{\cS} w(\br)\Big(\nabla u\cdot \br+\frac{1}{2!}\nabla^2 u:\br^2+\frac{1}{3!}\nabla^3 u\,\vdots\,\br^3+\cdots+\frac{1}{n!}\nabla^n u\cdot^{(n)}\br^n-u_{\br}\Big)^2 \,\ud V.
\end{align}
Let 
\begin{align}
S_{n-1}=u_{\br}-(\nabla u\cdot \br+\frac{1}{2!}\nabla^2 u:\br^2+\frac{1}{3!}\nabla^3 u\,\vdots\,\br^3+\cdots+\frac{1}{(n-1)!}\nabla^{n-1} u\cdot^{(n-1)}\br^{n-1}).
\end{align}
In the $n$-order Taylor series expansion of $u$, $S_{n-1}\approx \frac{1}{n!}\nabla^n u\cdot^{(n)}\br^n$.

On the other hand,
\begin{align}
S_{n}=S_{n-1}-\frac{1}{n!}\nabla^n u\cdot^{(n)}\br^n.
\end{align}
$E_n^{hg}$ can be simplified as
\begin{align}
\fF_n^{hg}&=\alpha\int_{\cS} w(\br) S_n^2 \ud V=\int_{\cS}w(\br) (S_{n-1}-\frac{1}{n!}\nabla^n u\cdot^{(n)}\br^n)^2 \ud V\notag\\
&=\alpha\int_{\cS}w(\br) \big(S_{n-1}^2+(\frac{1}{n!}\nabla^n u\cdot^{(n)}\br^n)^2-2 S_{n-1} (\frac{1}{n!}\nabla^n u\cdot^{(n)}\br^n) \big) \ud V\notag\\
&\approx\alpha\int_{\cS}w(\br) \big(S_{n-1}^2+(\frac{1}{n!}\nabla^n u\cdot^{(n)}\br^n)^2-2 (\frac{1}{n!}\nabla^n u\cdot^{(n)}\br^n)^2 \big) \ud V\notag\\
&=\alpha\int_{\cS}w(\br) \big(S_{n-1}^2-(\frac{1}{n!}\nabla^n u\cdot^{(n)}\br^n)^2\big) \ud V\notag\\
&=\fF_{n-1}^{hg}-\alpha(\frac{1}{n!}\nabla^n u\cdot^{(n)})^2\cdot^{(2 n)}\bK_{2 n}\notag.
\end{align}
Hence, $n$-order hourglass energy functional can be written as
\begin{align}
\fF^{hg}_n=\alpha\Big(\int_{\cS} w(\br) u_{\br}u_{\br} \,\ud V-(\nabla u)^2:\bK_2 -(\frac{1}{2!}\nabla^2 u)^2\cdot^{(4)}\bK_4 -\cdots-(\frac{1}{n!}\nabla^n u)^2 \cdot^{(2 n)}\bK_{2 n}\Big).
\end{align}
The $n$-order hourglass energy functional depends on $(1-n)$-order nonlocal operators, and the hourglass residual and hourglass stiffness matrix can be obtained with ease by the variation of the nonlocal operator.

\section{Variational principles based on the nonlocal operator}\label{sec:vp}
The problems based on variational principles start from the functional which describes the unknown functions defined in the domain and on the boundary. The residual is the gradient of the functional on the unknown vector, while the tangent stiffness matrix is the Hessian matrix of the functional on the unknown vector. The functional is usually expressed by the local operators such as divergence, curl and gradient. For simplicity, we assume the functional be a function on a  single local operator. The boundary terms can be handled the similar way.
%
%
%
Assuming four general functionals at a point which depend on the divergence of a vector field, the curl of a vector field, the gradient of a vector field and the gradient of a scalar field, respectively,
\begin{align}
\fF(\nabla\cdot\bv),\fF(\nabla \times\bv),\fF(\nabla \bv), \fF(\nabla v).
\end{align}
The examples are the strain energy functional in solid mechanics, the volume strain energy functional in solid mechanics, the wave vector form of electromagnetic field, and the thermal conduction, respectively.
%
%

\subsection{Divergence operator}
The first- and second-order derivatives of the functional $\fF(\nabla \cdot \bv)$ on operator $\nabla \cdot \bv$ are, respectively,
\begin{align}
p=\frac{\partial \fF(\nabla \cdot \bv)}{\partial (\nabla\cdot \bv)},\,
d=\frac{\partial p}{\partial (\nabla \cdot \bv)}=\frac{\partial^2 \fF(\nabla \cdot \bv)}{\partial (\nabla \cdot\bv)^T \partial (\nabla \cdot\bv)}
\end{align}
where $p$ is a scalar, $d$ is a scalar.

The residual and tangent stiffness matrix at a point are, respectively,
\begin{align}
\mathbf R_{div}=\tdelta \fF(\nabla \cdot \bv)&=\frac{\partial \fF(\nabla \cdot \bv)}{\partial (\nabla \cdot \bv)} \tnabla \cdot\tdelta \bv=p\, \tnabla \cdot\tdelta \bv\\
\mathbf K_{div}=\tdelta^2 \fF(\nabla \cdot \bv)&=(\tnabla \cdot\tdelta \bv)^T d\, \tnabla \cdot\tdelta \bv.
\end{align}
Let's consider the first variation of all particles, and let $p_{\bx}=\frac{\partial \fF(\nabla\cdot \bv_{\bx})}{\partial (\nabla\cdot \bv_{\bx})}$ .
\begin{align}
&\delta \fF(\nabla \cdot\bv)=\sum_{\Delta V_{\bx} \in \Omega}\Delta V_{\bx} \delta \fF_{\bx}=\sum_{\Delta V_{\bx} \in \Omega}\Delta V_{\bx}p_{\bx}\cdot (\nabla \cdot\delta\bv_{\bx})\notag\\
&=\sum_{\Delta V_{\bx} \in \Omega}\Delta V_{\bx}p_{\bx}\cdot\sum_{ \cS_{\bx}} w(\br) \Delta V_{\bx'} \bK_{\bx}^{-1} \br \cdot (\delta \bv_{\bx'}-\delta \bv_{\bx})\notag\\
&=\sum_{\Delta V_{\bx} \in \Omega}\Delta V_{\bx}\Big(-\sum_{ \cS_{\bx}} w(\br) \Delta V_{\bx'} \bK_{\bx}^{-1} \br \cdot \delta \bv_{\bx}\cdot p_{\bx}+\sum_{ \cS'_{\bx}} w(\br') \Delta V_{\bx'} \bK_{\bx'}^{-1} \br' \cdot \delta \bv_{\bx}\cdot p_{\bx'}\Big)\notag
\end{align}
In the second and third step, the dual-support is considered as follows. In the second step, the term with $\delta \bv_{\bx'}$ is the vector from $\bx$'s support, but is added to particle $\bx'$; since $\bx'\in \cS_{\bx}$, $\bx$ belongs to the dual-support $\cS'_{\bx'}$ of $\bx'$. In the third step, all the terms with $\delta \bv_{\bx}$ are collected from other particles whose supports contain $\bx$ and therefore form the dual-support of $\bx$. The terms with $\delta \bv_{\bx}$ in the first order variation $\delta \fF(\nabla \cdot \bv)=0$ are
\begin{align}
-\sum_{ \cS_{\bx}} w(\br) \Delta V_{\bx'} p_{\bx}\, \bK_{\bx}^{-1} \br+\sum_{ \cS'_{\bx}} w(\br') \Delta V_{\bx'} p_{\bx'}\, \bK_{\bx'}^{-1} \br'.
\end{align}
When any particle's volume $\Delta V_{\bx'}\to 0$, the continuous form is
\begin{align}
-\int_{\cS_{\bx}} w(\br) p_{\bx}\, \bK_{\bx}^{-1} \br \,\ud V_{\bx'}+\int_{\cS'_{\bx}} w(\br') p_{\bx'}\, \bK_{\bx'}^{-1} \br' \,\ud V_{\bx'}.\label{eq:sn2}
\end{align}
Eq.\ref{eq:sn2} is the nonlocal strong form for energy functional $\fF(\nabla \cdot \bv)$ with the corresponding local strong form is $-\nabla (d\,\nabla \cdot \bv)$. The local strong form can be obtained by integration by part of the energy functional in the whole domain.
Consider the variation of $\int_\Omega \fF(\nabla \cdot \bv_{\bx}) \ud V_{\bx}$
\begin{align}
&\delta\big( \int_\Omega \fF(\nabla \cdot\bv_{\bx}) \ud V_{\bx}\big)=\int_\Omega \delta \fF(\nabla\cdot \bv_{\bx}) \ud V_{\bx}\notag\\
&=\int_\Omega \pfrac{\fF(\nabla\cdot \bv_{\bx})}{(\nabla\cdot\bv_{\bx})}\cdot \nabla\cdot\delta \bv_{\bx} \ud V_{\bx}\notag\\
&=\int_\Omega p_\bx \nabla\cdot\delta \bv_\bx \ud V_{\bx}\notag\\
&=\int_{\partial\Omega} p_\bx\, \mathbf n_\bx \cdot\delta \bv_\bx \ud S_{\bx}\notag-\int_\Omega (\nabla p_\bx) \cdot\delta \bv_\bx \ud V_{\bx}\notag
\end{align}
For any point in $\Omega$, the term corresponding to $\bv_{\bx}$ is $-\nabla (d\, \nabla \cdot \bv_{\bx})$.

\subsection{Curl operator}
The first- and second-order derivatives of the functional $\fF(\nabla \times \bv)$ on operator $\nabla \times \bv$ are, respectively,
\begin{align}
\mathbf p=\frac{\partial \fF(\nabla \times \bv)}{\partial (\nabla \times\bv)},\,
\mathbf D=\frac{\partial\mathbf p}{\partial (\nabla \times \bv)}=\frac{\partial^2 \fF}{\partial (\nabla \times\bv)^T \partial (\nabla\times \bv)}.
\end{align}
In 3D $\mathbf p$ is a vector with length of 3,  $\mathbf D$ is $3\times 3$ matrices. 

The residual and tangent stiffness matrix for one point are
\begin{align}
\mathbf R_{curl}&=\tdelta \fF(\nabla \times \bv)=\frac{\partial \fF(\nabla \times \bv)}{\partial (\nabla \times \bv)} \tnabla \times \tdelta \bv=\mathbf p \tnabla \times \tdelta \bv\\\
\mathbf K_{curl}&=\tdelta^2 \fF(\nabla\times \bv)=(\tnabla \times \tdelta \bv)^T \bD \tnabla \times \tdelta \bv =(\tnabla \times \tdelta \bv)^T\bD \tnabla \times \tdelta \bv.
\end{align}

Let's consider the first variation of all particles, and let $\mathbf p_{\bx}=\frac{\partial \fF(\nabla \times \bv_{\bx})}{\partial (\nabla \times \bv_{\bx})}$.
\begin{align}
&\delta \fF(\nabla \times \bv)=\sum_{\Delta V_{\bx} \in \Omega}\Delta V_{\bx} \delta \fF_{\bx}=\sum_{\Delta V_{\bx} \in \Omega}\Delta V_{\bx}(\nabla \times \delta\bv_{\bx})\cdot \mathbf p_{\bx}\notag\\
&=\sum_{\Delta V_{\bx} \in \Omega}\Delta V_{\bx}\sum_{ \cS_{\bx}} w(\br) \Delta V_{\bx'} \bK_{\bx}^{-1} \br \times (\delta \bv_{\bx'}-\delta \bv_{\bx})\cdot \mathbf p_{\bx}\notag\\
&=\sum_{\Delta V_{\bx} \in \Omega}\Delta V_{\bx}\sum_{ \cS_{\bx}} w(\br) \Delta V_{\bx'} \mathbf p_{\bx} \times (\bK_{\bx}^{-1} \br) \cdot (\delta \bv_{\bx'}-\delta \bv_{\bx})\notag\\
&=\sum_{\Delta V_{\bx} \in \Omega}\Delta V_{\bx}\Big(-\sum_{ \cS_{\bx}} w(\br) \Delta V_{\bx'} \mathbf p_{\bx}\times(\bK_{\bx}^{-1} \br) \cdot \delta \bv_{\bx} +\sum_{ \cS'_{\bx}} w(\br') \Delta V_{\bx'} \mathbf p_{\bx'} \times (\bK_{\bx'}^{-1} \br') \cdot \delta \bv_{\bx}\Big).\notag
\end{align}
The relation $\mathbf a\cdot (\mathbf b \times \mathbf c)=\mathbf c\cdot (\mathbf a\times \mathbf b)=\mathbf b\cdot (\mathbf c \times \mathbf a)$ is used in the third step. In the third and fourth step, the dual-support is considered as follows. In the third step, the term with $\delta \bv_{\bx'}$ is the vector from $\bx$'s support, but is added to particle $\bx'$; since $\bx'\in \cS_{\bx}$, $\bx$ belongs to the dual-support $\cS'_{\bx'}$ of $\bx'$. In the fourth step, all the terms with $\delta \bv_{\bx}$ are collected from other particles whose supports contain $\bx$ and therefore form the dual-support of $\bx$. The terms with $\delta \bv_{\bx}$ in the first order variation $\delta \fF(\nabla \times \bv)=0$ are
\begin{align}
-\sum_{ \cS_{\bx}} w(\br) \Delta V_{\bx'} \mathbf p_{\bx}\times(\bK_{\bx}^{-1} \br) +\sum_{ \cS'_{\bx}} w(\br') \Delta V_{\bx'} \mathbf p_{\bx'} \times (\bK_{\bx'}^{-1} \br').
\end{align}
When any particle's volume $\Delta V_{\bx'}\to 0$, the continuous form is
\begin{align}
-\int_{ \cS_{\bx}} w(\br) \mathbf p_{\bx}\times(\bK_{\bx}^{-1} \br) \,\ud V_{\bx'} +\int_{ \cS'_{\bx}} w(\br') \mathbf p_{\bx'} \times (\bK_{\bx'}^{-1} \br') \,\ud V_{\bx'}.\label{eq:sn3}
\end{align}
Eq.\ref{eq:sn3} is the strong form for energy functional $\fF(\nabla \times\bv)$, where the corresponding local strong from obtained by integration by part of the energy functional is $-\nabla \times (\bD \nabla \times \bv)$, which is obtained as follows.
In order to derive the local strong form of functional $\fF(\nabla \times \bv)$, consider the variation of $\int_\Omega \fF(\nabla \times \bv_{\bx}) \ud V_{\bx}$
\begin{align}
&\delta\big( \int_\Omega \fF(\nabla \times\bv_{\bx}) \ud V_{\bx}\big)=\int_\Omega \delta \fF(\nabla\times \bv_{\bx}) \ud V_{\bx}\notag\\
&=\int_\Omega \pfrac{\fF(\nabla\times \bv_{\bx})}{(\nabla\times\bv_{\bx})}\cdot \nabla\times\delta \bv_{\bx} \ud V_{\bx}\notag\\
&=\int_\Omega \mathbf p_\bx \cdot\nabla\times\delta \bv_\bx \ud V_{\bx}\notag\\
&=\int_{\partial\Omega} \mathbf p_\bx\times \mathbf n_\bx \cdot\delta \bv_\bx \ud S_{\bx}\notag-\int_\Omega (\nabla \times\mathbf p_\bx) \cdot\delta \bv_\bx \ud V_{\bx}\notag
\end{align}
For any point in $\Omega$, the term corresponding to $\delta\bv_{\bx}$ is $-\nabla \times (\mathbf D \nabla \times \bv_{\bx})$.

\subsection{Gradient operator of vector field}
The first- and second-order derivatives of the functional $\fF(\nabla \bv)$ on the operator $\nabla \bv$ are, respectively,
\begin{align}
\bS=\frac{\partial \fF(\nabla \bv)}{\partial (\nabla \bv)},\,
\mathbf D=\frac{\partial\bS}{\partial (\nabla \bv)}=\frac{\partial^2 \fF(\nabla \bv)}{\partial (\nabla \bv)^T \partial (\nabla \bv)}
\end{align}
$\bS$ is a $3\times 3$ tensor. When $\nabla \bv$ is the deformation gradient with respect to the initial configuration, $\bS$ is the first Piola-Kirchhoff stress. $\mathbf D$ is a $3\times 3\times 3\times 3$ tensor. The fourth-order tensor $\mathbf D$ can be flattened into a $9\times 9$ matrix as long as the $\nabla \bv$ is flattened into a vector with length of 9. When $\nabla \bv$ is the deformation gradient with respect to the initial configuration, $\bD$ is the material tensor in solid mechanics.

The residual and stiffness matrix at one point are, respectively,
\begin{align}
\mathbf R_{grad}&=\tdelta \fF(\nabla \bv)=\frac{\partial \fF(\nabla \bv)}{\partial (\nabla \bv)} \tnabla \tdelta \bv=\bS\tnabla \tdelta \bv\\
\mathbf K_{grad}&=\tdelta^2 \fF(\nabla \bv)=(\tnabla \tdelta \bv)^T \bD \tnabla \tdelta \bv= (\tnabla \tdelta \bv)^T \bD\tnabla \tdelta \bv.
\end{align} 
Let's consider the first variation of all particles, and let $\bS_{\bx}=\frac{\partial \fF(\nabla \bv_{\bx})}{\partial (\nabla \bv_{\bx})}$.
\begin{align}
&\delta \fF(\nabla \bv)=\sum_{\Delta V_{\bx} \in \Omega}\Delta V_{\bx} \delta \fF_{\bx}=\sum_{\Delta V_{\bx} \in \Omega}\Delta V_{\bx} \bS_{\bx}\cdot (\nabla \delta\bv_{\bx})\notag\\
&=\sum_{\Delta V_{\bx} \in \Omega}\Delta V_{\bx}\bS_{\bx}\cdot\sum_{ \cS_{\bx}} w(\br) \Delta V_{\bx'} \bK_{\bx}^{-1} \br \otimes (\delta \bv_{\bx'}-\delta \bv_{\bx})\notag\\
&=\sum_{\Delta V_{\bx} \in \Omega}\Delta V_{\bx}\Big(-\sum_{ \cS_{\bx}} w(\br) \Delta V_{\bx'} \bK_{\bx}^{-1} \br \otimes \delta \bv_{\bx}\cdot \bS_{\bx}+\sum_{ \cS'_{\bx}} w(\br') \Delta V_{\bx'} \bK_{\bx'}^{-1} \br' \otimes \delta \bv_{\bx}\cdot \bS_{\bx'}\Big).\notag
\end{align}
In the second and third step, the dual-support is considered as follows. In the second step, the term with $\delta \bv_{\bx'}$ is the vector from $\bx$'s support, but is added to particle $\bx'$; since $\bx'\in \cS_{\bx}$, $\bx$ belongs to the dual-support $\cS'_{\bx'}$ of $\bx'$. In the third step, all the terms with $\delta \bv_{\bx}$ are collected from other particles whose supports contain $\bx$ and therefore form the dual-support of $\bx$. The terms with $\delta \bv_{\bx}$ in the first order variation $\delta \fF(\bv)=0$ are
\begin{align}
-\sum_{ \cS_{\bx}} w(\br) \Delta V_{\bx'} \bS_{\bx}\cdot \bK_{\bx}^{-1} \br+\sum_{ \cS'_{\bx}} w(\br') \Delta V_{\bx'} \bS_{\bx'}\cdot \bK_{\bx'}^{-1} \br'.
\end{align}
When any particle's volume $\Delta V_{\bx'}\to 0$, the continuous form is
\begin{align}
-\int_{\cS_{\bx}} w(\br) \bS_{\bx}\cdot \bK_{\bx}^{-1} \br \,\ud V_{\bx'}+\int_{\cS'_{\bx}} w(\br') \bS_{\bx'}\cdot \bK_{\bx'}^{-1} \br' \,\ud V_{\bx'}.\label{eq:sn1}
\end{align}
Eq.\ref{eq:sn1} is the nonlocal strong form of energy functional $\fF_3$, where the local strong form obtained by integration by part of the energy functional is $-\nabla \cdot \bS_{\bx}$.
If $\nabla \bv$ denotes the deformation gradient with respect to the initial configuration and $\fF(\nabla \bv)$ is the strain energy density, $\bS_{\bx}$ is the first Piola-Kirchhoff stress and the Eq.\ref{eq:sn1} is the key expression in the dual-horizon peridynamics \cite{Ren2015,ren2017dual}. 

The local strong form $-\nabla \cdot \bS_\bx$ is obtained as follows.
In order to derive the local strong form of functional $\fF(\nabla \bv)$, consider the variation of $\int_\Omega \fF(\nabla \bv_{\bx}) \ud V_{\bx}$
\begin{align}
&\delta\big( \int_\Omega \fF(\nabla \bv_{\bx}) \ud V_{\bx}\big)=\int_\Omega \delta \fF(\nabla \bv_{\bx}) \ud V_{\bx}\notag\\
&=\int_\Omega \pfrac{\fF(\nabla \bv_{\bx})}{(\nabla\bv_{\bx})}\cdot \nabla\delta \bv_{\bx} \ud V_{\bx}\notag\\
&=\int_\Omega \bS_\bx \cdot\nabla\delta \bv_\bx \ud V_{\bx}\notag\\
&=\int_{\partial\Omega} \bS_\bx\cdot \mathbf n_\bx \cdot\delta \bv_\bx \ud S_{\bx}\notag-\int_\Omega (\nabla \cdot\bS_\bx) \cdot\delta \bv_\bx \ud V_{\bx}\notag
\end{align}
For any point in $\Omega$, the term corresponding to $\delta\bv_{\bx}$ is $-\nabla \cdot \bS_{\bx}$.

\subsection{Gradient operator of scalar field}
The first- and second-order derivatives of the functional $\fF(\nabla v)$ on the operator $\nabla v$ are, respectively,
\begin{align}
\mathbf p=\frac{\partial \fF(\nabla v)}{\partial (\nabla v)},\,
\mathbf D=\frac{\partial\mathbf p}{\partial (\nabla v)}=\frac{\partial^2 \fF(\nabla v)}{\partial (\nabla v)^T \partial (\nabla v)}.
\end{align}
In 3D, $\mathbf p$ is a 3-vector. $\mathbf D$ are $3\times 3$ matrices. 

The residual and tangent stiffness matrix at one point are, respectively,
\begin{align}
\mathbf R_{grad}&=\tdelta \fF(\nabla v)=\frac{\partial \fF(\nabla v)}{\partial (\nabla v)} \tnabla \tdelta v=\mathbf p \tnabla \tdelta v\\
\mathbf K_{grad}&=\tdelta^2 \fF(\nabla v)=(\tnabla \tdelta v)^T \bD \tnabla \tdelta v.
\end{align} 
In order to derive the nonlocal strong form, let's consider the first variation of all particles, and let $\mathbf p_{\bx}=\frac{\partial \fF(\nabla v_{\bx})}{\partial (\nabla v_{\bx})}$.
\begin{align}
&\delta \fF(\nabla v)=\sum_{\Delta V_{\bx} \in \Omega}\Delta V_{\bx} \delta \fF_{\bx}=\sum_{\Delta V_{\bx} \in \Omega}\Delta V_{\bx}\nabla \delta v_{\bx}\cdot \bS_{\bx}\notag\\
&=\sum_{\Delta V_{\bx} \in \Omega}\Delta V_{\bx}\Big(\sum_{ \cS_{\bx}} w(\br) \Delta V_{\bx'} \bK_{\bx}^{-1} \br (\delta v_{\bx'}-\delta v_{\bx})\cdot \bS_{\bx}\Big)\notag\\
&=\sum_{\Delta V_{\bx} \in \Omega}\Delta V_{\bx}\Big(-\sum_{ \cS_{\bx}} w(\br) \Delta V_{\bx'} \bK_{\bx}^{-1} \br \delta v_{\bx}\cdot \bS_{\bx}+\sum_{ \cS'_{\bx}} w(\br') \Delta V_{\bx'} \bK_{\bx'}^{-1} \br' \delta v_{\bx}\cdot \bS_{\bx'}\Big)\notag
\end{align}
In the second and third step, the dual-support is considered as follows. In the second step, the term with $\delta v_{\bx'}$ is the vector from $\bx$'s support, but is added to particle $\bx'$; since $\bx'\in \cS_{\bx}$, $\bx$ belongs to the dual-support $\cS'_{\bx'}$ of $\bx'$. In the third step, all the terms with $\delta v_{\bx}$ are collected from other particles whose supports contain $\bx$ and therefore form the dual-support of $\bx$. The terms with $\delta v_{\bx}$ in the first order variation $\delta \fF(\nabla v)=0$ are
\begin{align}
-\sum_{ \cS_{\bx}} w(\br) \Delta V_{\bx'} \bK_{\bx}^{-1} \br\cdot \bS_{\bx}+\sum_{ \cS'_{\bx}} w(\br') \Delta V_{\bx'} \bK_{\bx'}^{-1} \br'\cdot\bS_{\bx'}.
\end{align}
When any particle's volume $\Delta V_{\bx'}\to 0$, the continuous form is
\begin{align}
-\int_{\cS_{\bx}} w(\br) \bK_{\bx}^{-1} \br\cdot\bS_{\bx} \,\ud V_{\bx'}+\int_{\cS'_{\bx}} w(\br') \bK_{\bx'}^{-1} \br'\cdot\bS_{\bx'} \,\ud V_{\bx'}.\label{eq:sn4}
\end{align}
Eq.\ref{eq:sn4} is the nonlocal strong form for energy functional $\fF(\nabla v)$.

The simplest example for this energy functional is
\begin{align}
\fF(\nabla T)=\frac{1}{2}\kappa \nabla T \cdot\nabla T\notag,
\end{align}
where $T$ is the temperature, $\kappa$ is the thermal conductivity. The local strong form corresponding to Eq.\ref{eq:sn4} is $-\nabla\cdot \mathbf p$, where $\mathbf p=\kappa\nabla T$.

The local strong form $-\nabla \cdot \mathbf p_\bx$ is obtained as follows.
In order to derive the local strong form of functional $\fF(\nabla T)$, consider the variation of $\int_\Omega \fF(\nabla T) \ud V_{\bx}$
\begin{align}
&\delta\big( \int_\Omega \frac{1}{2}\kappa \nabla T_\bx \cdot\nabla T_\bx \ud V_{\bx}\big)\notag\\
&=\int_\Omega \kappa \nabla T_\bx\cdot \nabla\delta T_{\bx} \ud V_{\bx}\notag\\
&=\int_{\partial\Omega} \kappa \nabla T_\bx\cdot \mathbf n_\bx \cdot\delta T_\bx \ud S_{\bx}\notag-\int_\Omega (\nabla \cdot\kappa \nabla T_\bx) \cdot\delta T_\bx \ud V_{\bx}\notag
\end{align}
For any point in $\Omega$, the term corresponding to $\delta T_{\bx}$ is $-\nabla \cdot (\kappa \nabla T_\bx)$.

\section{Applications}\label{sec:appli}
\subsection{One dimensional beam and bar test}
The energy functionals of cantilever beam and bar are, respectively,
\begin{align}
\fF_{beam}(u)=\frac{1}{2}\int_0^L (u_{,xx} EI u_{,xx} -u q) dx\label{eq:fbe}\\
\fF_{bar}(u)=\frac{1}{2}\int_0^L (u_{,x} EA u_{,x} -u q) dx.\label{eq:fba}
\end{align}
We consider the boundary conditions in Eq.\ref{eq:beamBC} of the cantilever beam with a concentrated transverse load $P=1$ is applied on the end.
\begin{align}
u(0)=0, \frac{ \ud u}{ \ud x}|_{x=0}=0.\label{eq:beamBC} 
\end{align}
For the uniform bar, the left side is fixed and the other side is applied with a load $P=1$.
The theoretical solution for beam and bar are, respectively,
\begin{align}
u(x)=\frac{P}{6 E I}(3 L x^2-x^3), u(x)=\frac{P x}{EA},
\end{align}
where $EI=1, EA=1$ are the stiffness coefficient, $L=1$ is the length of the beam.
The residual and tangent stiffness matrix of Eq.\ref{eq:fbe} and Eq.\ref{eq:fba} are obtained by simply replacing $\tdelta u_{,x}$ and $\tdelta u_{,xx}$ with Eq.\ref{eq:FGdef} and Eq.\ref{eq:avsd} in the first and second variation of the energy functional, respectively.
The L2-norm is calculated by
\begin{align}
\|\bu\|_{L2}=\sqrt{\frac{\sum_j (\bu_j-\bu_j^{exact})\cdot (\bu_j-\bu_j^{exact}) \Delta V_j}{\sum_j \bu^{exact}_j\cdot \bu_j^{exact} \Delta V_j}}.\label{eq:uerror}
\end{align}
The convergence of the L2-norm for the displacement of bar under tension is shown in Fig.\ref{fig:barL2}. The convergence of the L2-norm for the deflection of cantilever beam is shown in Fig.\ref{fig:beamL2}. With the refinement in discretization, the numerical results converge to the theoretical solutions at a rate $r\approx 1$.
\begin{figure}
	\centering
		\includegraphics[width=9cm]{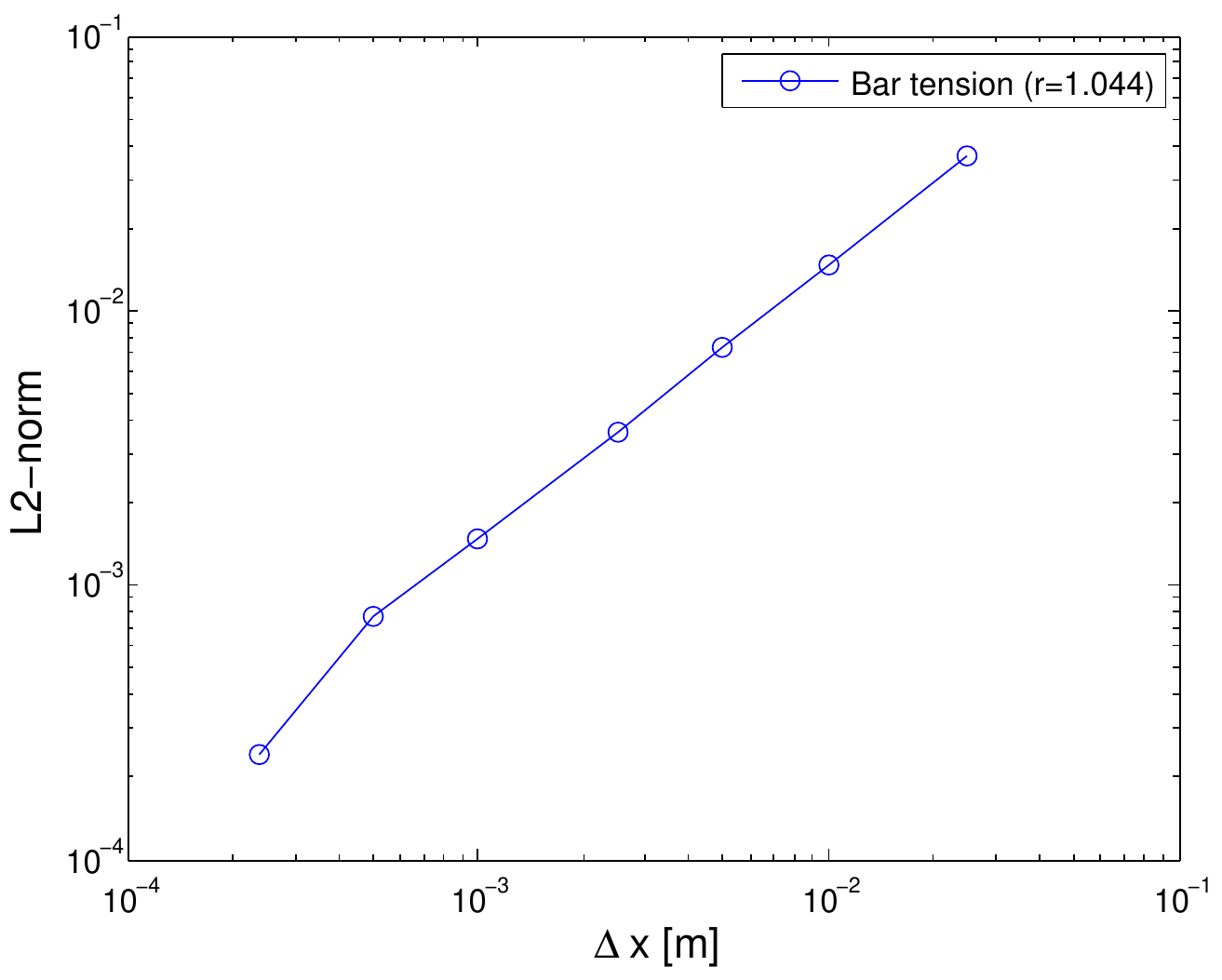}
	\caption{Convergence of the L2-norm for the displacement of bar under tension.}
	\label{fig:barL2}
\end{figure} 
\begin{figure}
	\centering
		\includegraphics[width=9cm]{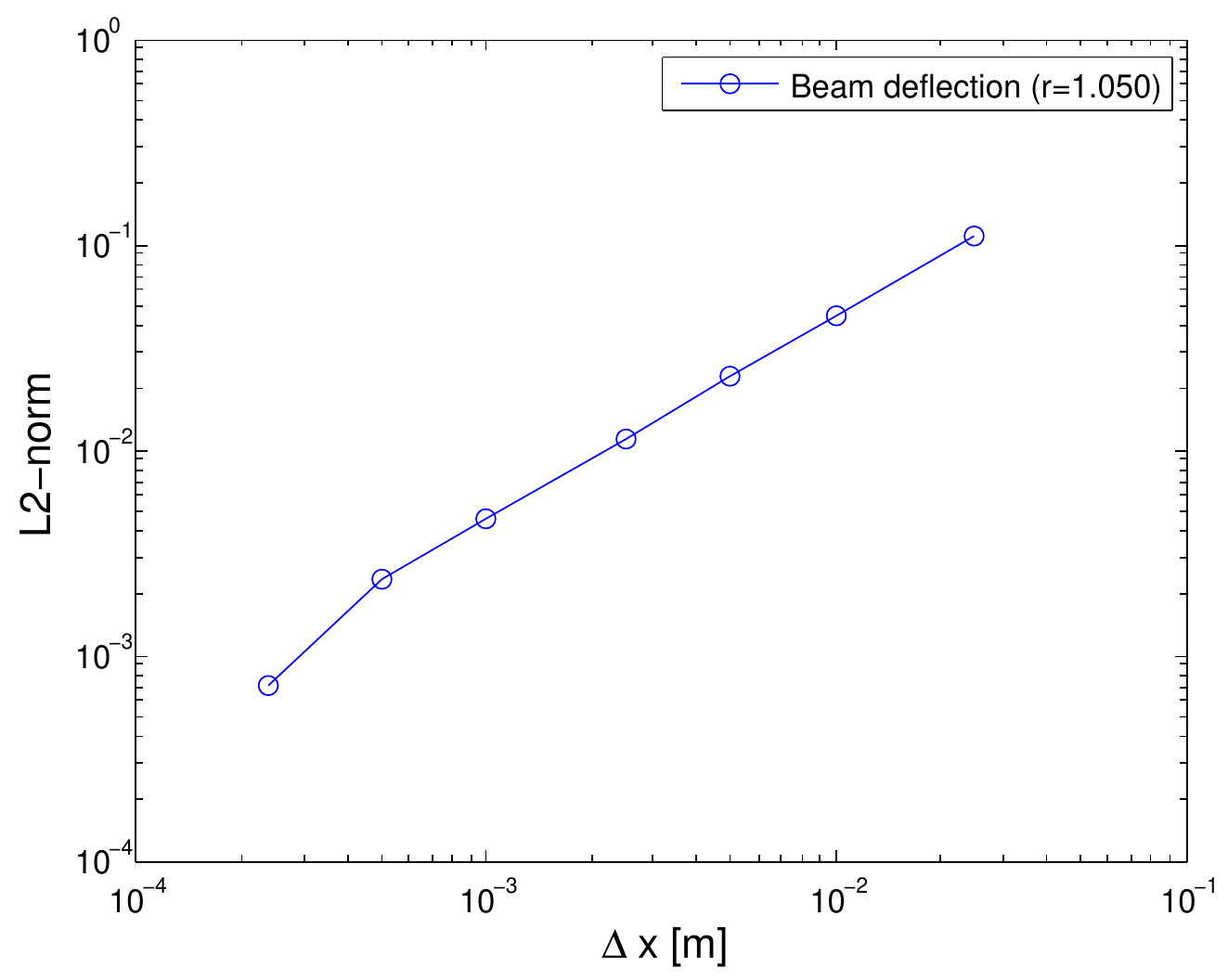}
	\caption{Convergence of the L2-norm for the deflection of cantilever beam.}
	\label{fig:beamL2}
\end{figure}
\subsection{1D Schr\"{o}dinger equation}
In this section, we test the accuracy of the eigenvalue problem based on the nonlocal operator. The Schr\"{o}dinger equation written in adimensional units for a one-dimensional harmonic oscillator is
\begin{align}
\big[-\frac{1}{2} \frac{\partial^2}{\partial x^2}+V(x)\big] \phi(x)=\lambda \phi(x),\quad V(x)=\frac{1}{2}\omega^2 x^2.
\end{align}
For simplicity, we use $\omega=1$. The particles are distributed with constant or variables spacing $\Delta x$ on the region [-10,10].

The exact wave functions and eigenvalues can be expressed as
\begin{align}
\phi_n(x)=H_n(x) \exp(\pm \frac{x^2}{2}), \quad \lambda_n=n+\frac{1}{2},
\end{align} 
where $n$ is a non-negative integer. $H_n(x)$ is the $n$-order Hermite polynomial.
We calculate the lowest eigenvalue and compare the numerical result with $\lambda_0=0.5$. The convergence plot of the error is shown in Fig.\ref{fig:scheig}. With the decrease of grid spacing, the numerical result converges to the exact result at a rate of $r\approx 2$.
\begin{figure}
	\centering
		\includegraphics[width=9cm]{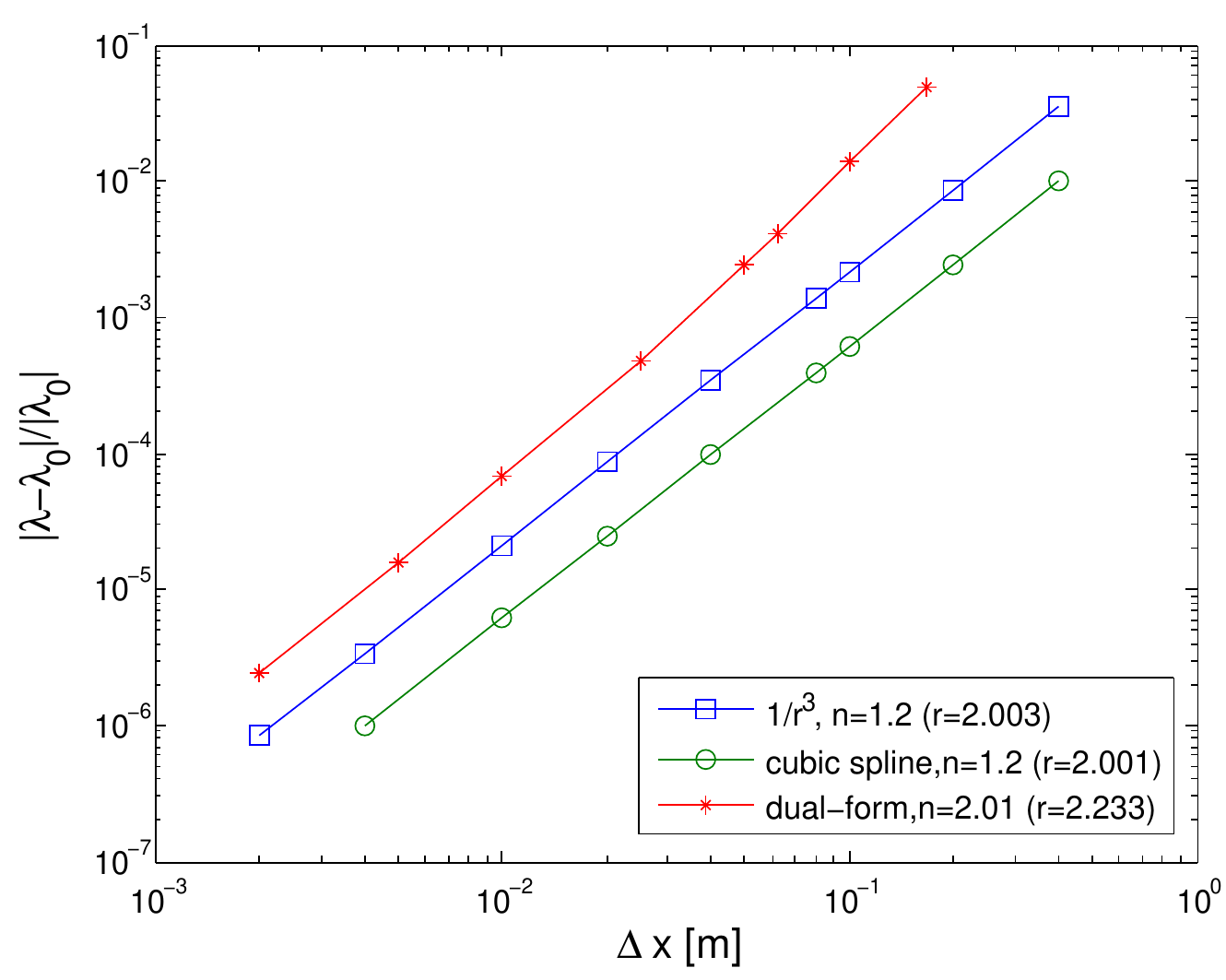}
	\caption{Convergence of the lowest eigenvalue for a one-dimensional harmonic oscillator; $1/r^3$ is the weight function; support radius is selected as $h=n\Delta x$; dual-form with influence function $1/r^3$ uses an inhomogenous discretization in Fig.\ref{fig:schdualform}; {the particle spacing in dual-form is selected as the minimal particle spacing in the discretization.}}
	\label{fig:scheig}
\end{figure} 
The discretization of the dual-form is given in Fig.\ref{fig:schdualform}. The first three wave functions are given in Fig.\ref{fig:schwave}.
\begin{figure}
	\centering
		\includegraphics[width=9cm]{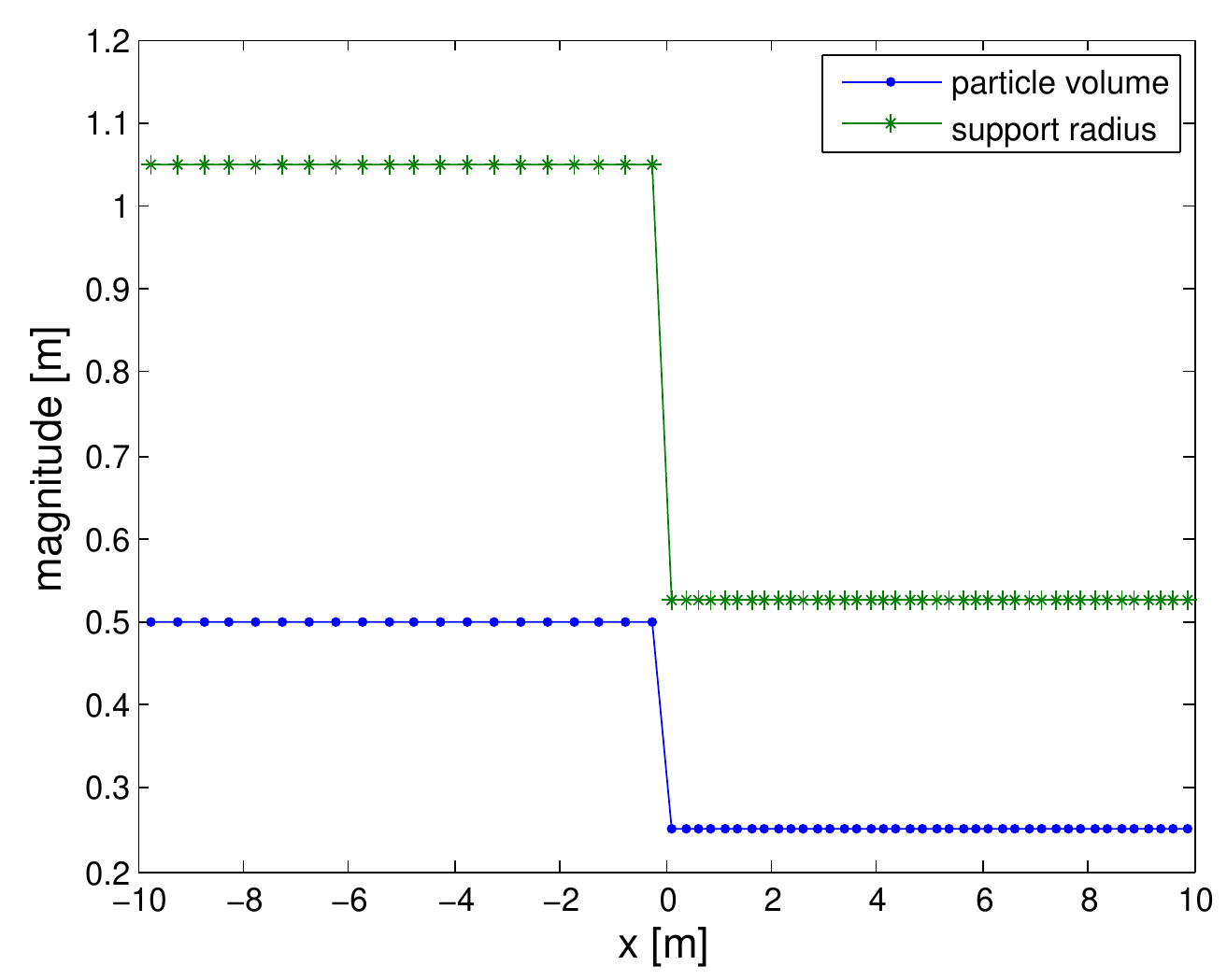}
	\caption{Discretization of the dual form based on inhomogeneous discretization.}
	\label{fig:schdualform}
\end{figure} 
\begin{figure}
	\centering
		\includegraphics[width=9cm]{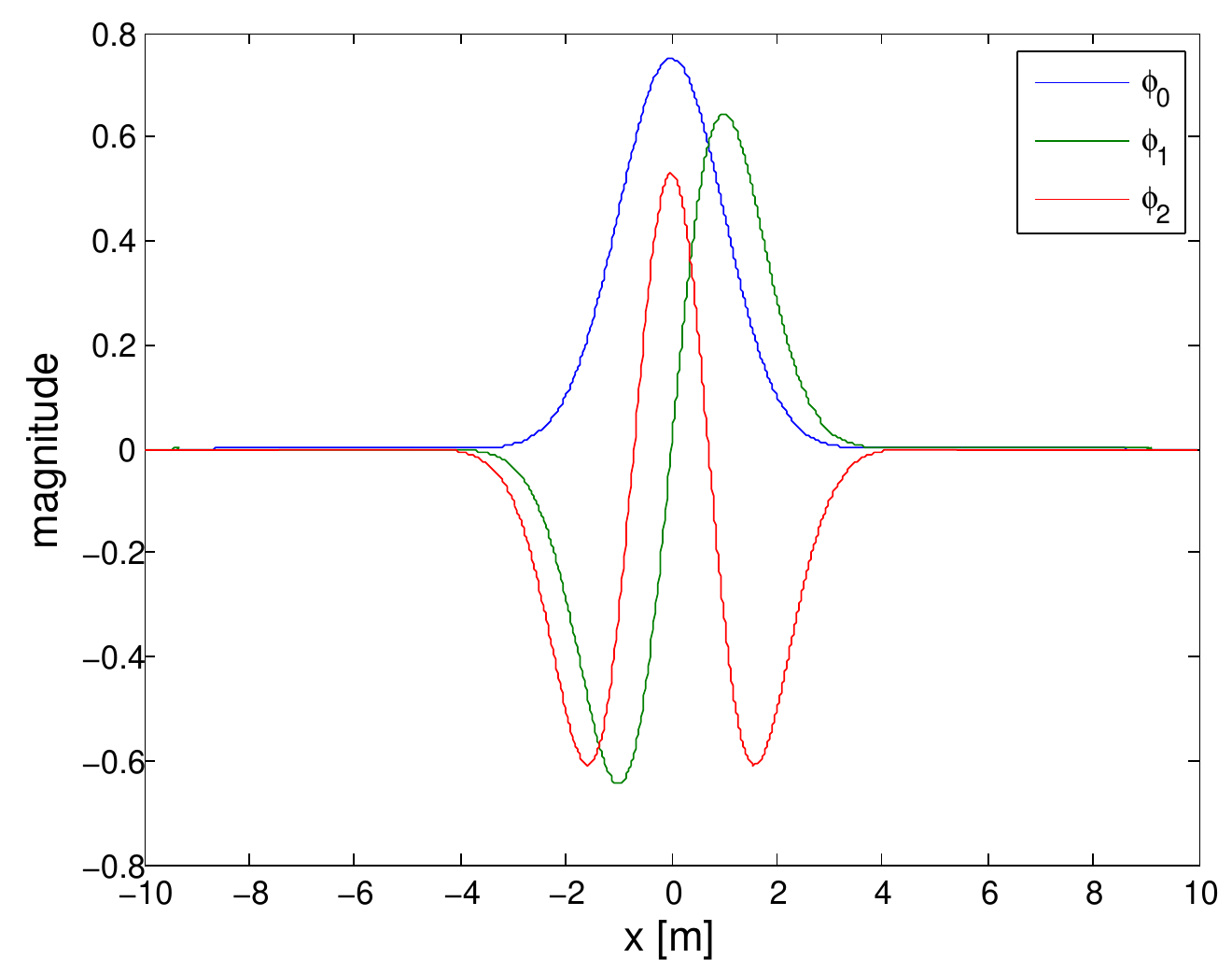}
	\caption{First three wave functions.}
	\label{fig:schwave}
\end{figure}
\subsection{Poisson equation}
In this section, we test the Poisson equation 
\begin{align}
\nabla^2 u=f(x,y),\quad (x,y)\in (0,1)\times (0,1),
\end{align}
where $f(x,y)=2 x (y-1)(y-2 x+x y+2) e^{x-y}$, and the boundary conditions
\begin{align}
u(x,0)=u(x,1)=0,\, x\in [0,1]\notag\\
u(0,y)=u(1,y)=0,\, y\in [0,1].\notag
\end{align}
The analytic solution is 
\begin{align}
u(x,y)=x(1-x)y(1-y)e^{x-y}.
\end{align}

The corresponded energy functional is
\begin{align}
\fF=\int_\Omega \big(-\frac{1}{2}\nabla u\cdot \nabla u -f(x,y) u\big)d\Omega.
\end{align}
The first and second variation of $\Pi$ lead to the global residual and stiffness matrix
\begin{align}
\mathbf R_g&=\sum_{\Delta V_i\in\Omega} \Delta V_i\big(-\tnabla u \cdot \tnabla \tdelta u-f(x,y)\tdelta u\big) \\
\mathbf K_g&=\sum_{\Delta V_i\in\Omega} \Delta V_i\big(-\tnabla \tdelta u \cdot \tnabla \tdelta u\big) 
\end{align}
The support radius is selected as $h=1.2 \Delta \bx$. We test the convergence of the L2 error for $u$ field under difference discretizations. The convergent plot is given in Fig.\ref{fig:P2dL2} with convergence rate of $r=0.9567$. The contours of $u$ field with and without hourglass control are shown in Fig.\ref{fig:P2contour}. It can be seen that the hourglass control can stabilize the solution.
\begin{figure}
	\centering
		\includegraphics[width=9cm]{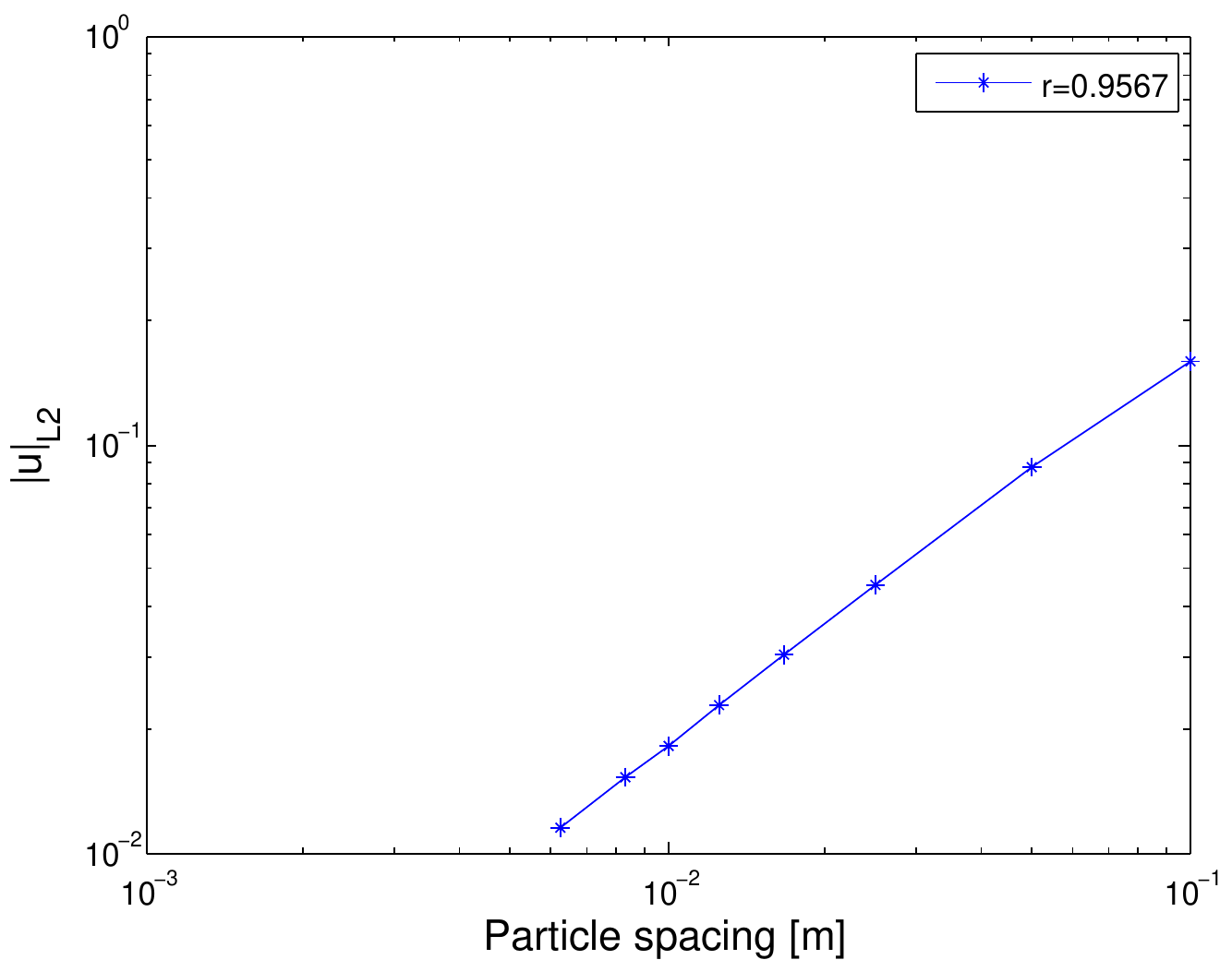}
	\caption{Convergence of the L2 error of the displacement.}
	\label{fig:P2dL2}
\end{figure}

\begin{figure}
 \centering
 \subfigure[]{
 \label{fig:P2hg}
 \includegraphics[width=.4\textwidth]{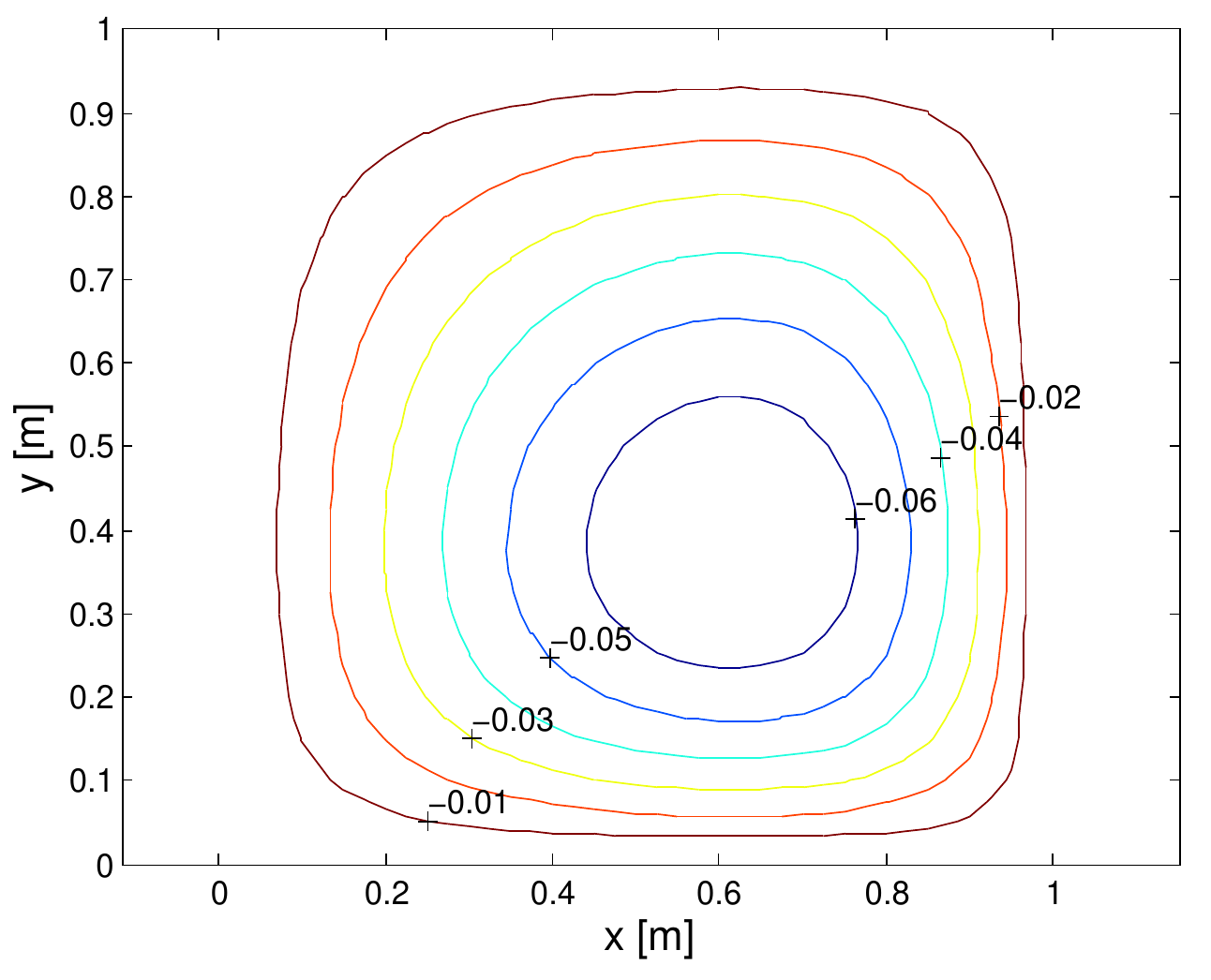}}
 \vspace{.1in}
 \subfigure[]{
 \label{fig:P2noHg}
 \includegraphics[width=.4\textwidth]{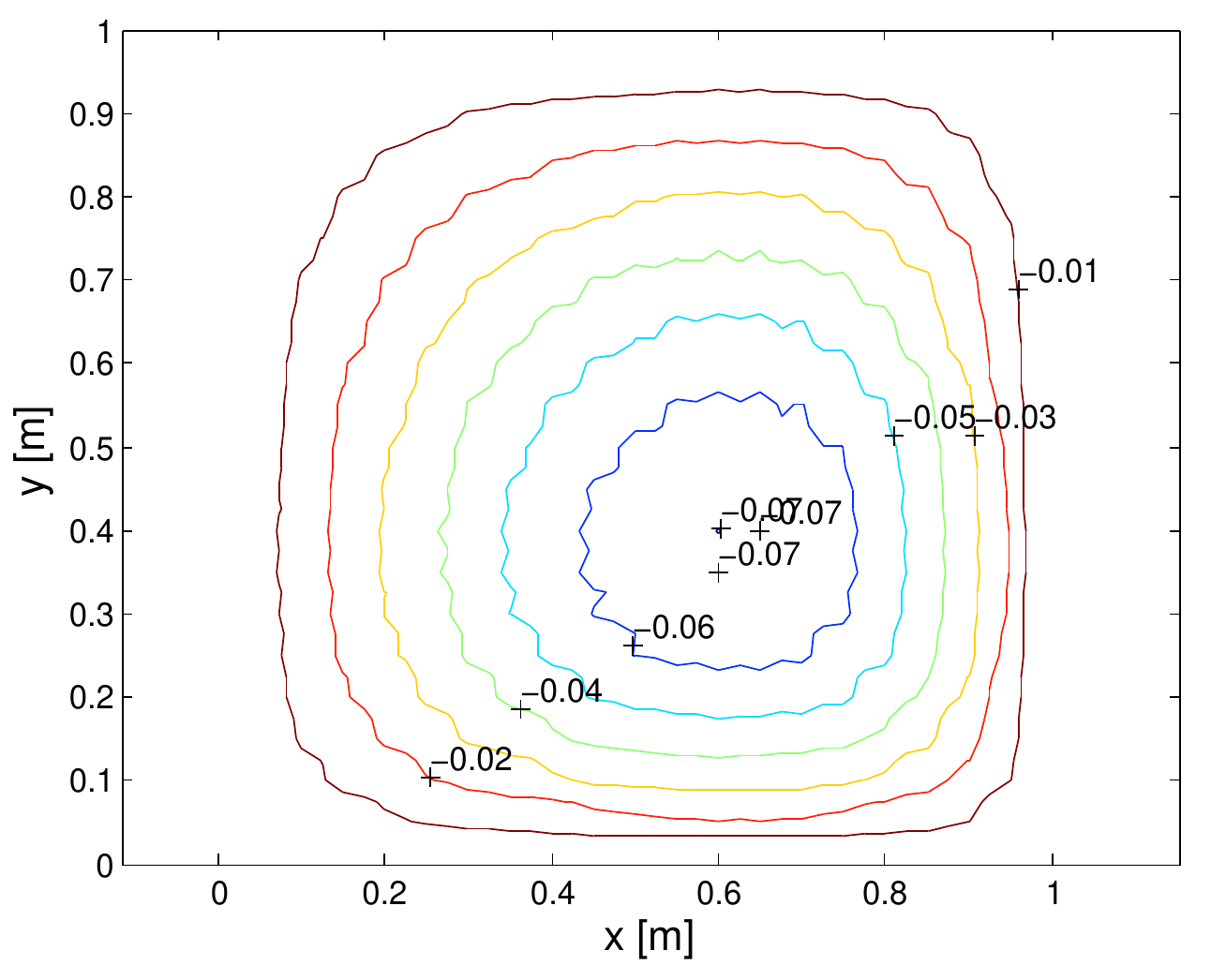}}\\
 \vspace{.3in}
\caption{Contour of $u$ with hourglass control $\mu=0.1$ and without hourglass control for discretization $40\times 40$.}
\label{fig:P2contour}
\end{figure} 

\subsection{Nonlocal theory for linear small strain elasticity}
The elastic energy of a body $V$ is given by the quadratic functional \cite{polizzotto2001nonlocal,bavzant2002nonlocal}
\begin{align}
\fF=\frac{1}{2} \int_V\int_V \bvps^T(\bx) \bD_e(\bx,\bx') \bvps(\bx') \,\ud \bx'd \bx=\frac{1}{2} \int_V \bvps^T(\bx) \bsig(\bx) \,\ud \bx\label{eq:nonlocalelastic},
\end{align}
where $\bvps(\bx)$=strain field, $\bD_e(\bx,\bx')$=generalized form of the elastic stiffness and
\begin{align}
\bsig(\bx)=\int_V \bD_e(\bx,\bx') \bvps(\bx') \,\ud \bx'
\end{align}
is the stress dependent on the strain field in $V$.
Only if $\bD_e(\bx,\bx')=\bD_e(\bx)\delta(\bx-\bx')$, Eq.\ref{eq:nonlocalelastic} reduces to
\begin{align}
\fF=\frac{1}{2} \int_V \bvps^T(\bx) \bD_e(\bx) \bvps(\bx)d \bx=\int_V \fF[\bvps(\bx),\bx]d \bx\label{eq:localelastic},
\end{align}
where $\fF(\bvps,\bx)=\frac{1}{2}\bvps^T \bD_e(\bx) \bvps$.

{It can be assumed that} the interaction effects decay with distance between the two points $\bx$ and $\bx'$, i.e.,
\begin{align}
\bD_e(\bx,\bx')=\bD_e \alpha(\bx,\bx'),
\end{align}
where $\alpha$ is certain attenuation function satisfying the normalizing condition
\begin{align}
\int_{V} \alpha(\bx,\bx') \,\ud \bx'=1\label{eq:alphaN}.
\end{align}
$\alpha$ is also called the nonlocal weight function or the nonlocal averaging function, and is often assumed to have the form of Gauss distribution function
\begin{align}
\alpha_\infty (r)=( l \sqrt{2 \pi})^{-N_{dim}} \exp(-\frac{r^2}{2 l^2}),
\end{align}
where $l$ is the parameter with the dimension of length, $N_{dim}$ the number of spatial dimensions. For reasons of computational efficiency, the attenuation function is often selected as the finite support, e.g., the polynomial bell-shaped function,
\begin{align}
\alpha(\br)=c \big(\max(0, 1-\frac{r^2}{R^2})\big)^2,
\end{align}
where $c$ is determined by the normalizing condition Eq.\ref{eq:alphaN}.

Then the stress-strain law reads
\begin{align}
\bsig(\bx)=\int_V \bD_e \alpha(\bx,\bx') \bvps(\bx') \,\ud \bx'=\bD_e\int_V \alpha(\bx,\bx') \bvps(\bx') \,\ud \bx'=\bD_e \bar{\bvps}(\bx),
\end{align}
where 
\begin{align}
\bar{\bvps}(\bx)=\int_V \alpha(\bx,\bx') \bvps(\bx') \,\ud \bx'
\end{align}
is the nonlocal strain.

With the aid of nonlocal operator and its variation,
\begin{align}
\bvps&=\frac{1}{2}\big(\tnabla \bu+(\tnabla \bu)^T\big)\notag\\
\tdelta \bvps&=\frac{1}{2}\big(\tnabla \tdelta\bu+(\tnabla \tdelta\bu)^T\big),\notag
\end{align}
where $\tnabla \tdelta \bu$ {is the nonlocal gradient operator in Eq.}\ref{eq:FGdef}, the residual and tangent stiffness matrix of nonlocal energy functional Eq.\ref{eq:nonlocalelastic} are
\begin{align}
\mathbf R_g&=\tdelta \fF=\sum_{\bx \in V}\sum_{\bx'\in V} \bvps^T(\bx) \bD_e(\bx,\bx') \tdelta\bvps(\bx') \Delta V_{\bx'}\Delta V_{\bx}\\
\mathbf K_g&=\tdelta^2 \fF=\sum_{\bx \in V}\sum_{\bx'\in V} \tdelta\bvps^T(\bx) \bD_e(\bx,\bx') \tdelta\bvps(\bx') \Delta V_{\bx'}\Delta V_{\bx}.
\end{align}
It is found that the tangent stiffness matrix for nonlocal elasticity is equivalent to the matrix multiplication on the variational form of the nonlocal operator on each particle. The Neumann boundary conditions and Dirichlet boundary conditions can be applied directly on the residual and stiffness matrix.
\subsection{Nonhomogeneous biharmonic equation}
This example tests the nonlocal Hessian operator. The nonhomogeneous biharmonic equation reads
\begin{align}
\nabla^2 \nabla^2 w=q_0,\quad (x,y)\in (0,1)\times (-1/2,1/2)\label{eq:nbh}
\end{align}
with boundary conditions
\begin{align}
w(x,-1/2)=w(x,1/2)=0,\,x\in [0,1]\notag\\
w(0,y)=w(1,y)=0,\,y\in [-1/2,1/2]\notag.
\end{align}
This biharmonic equation corresponds to the simply support square plate subjected to uniform load with parameters such as length $a=1$ m, thickness $t=0.01$ m, uniform pressure $q_0$=-100 N, Poisson ratio $\nu=0$, elastic modulus $E=30$ GPa and $D_0=\frac{E t^3}{12(1-\nu^2)}$.

The analytic solution for this plate is denoted by \cite{timoshenko1959theory}
\begin{align}
w=\frac{4 q_0 a^4}{\pi^5 D_0}\sum_{m=1,3,\cdots}^{\infty} \frac{1}{m^5}\Big(1-\frac{\alpha_m \tanh \alpha_m+2}{2 \cosh \alpha_m}\cosh\frac{2 \alpha_m y}{a}+\frac{\alpha_m}{2\cosh\alpha_m}\frac{2y}{a}\sinh\frac{2\alpha_m y}{a}\Big)\sin\frac{m \pi x}{a},
\end{align}
where $\alpha_m=\frac{m \pi}{2}$. 

The equivalent energy functional of Eq.\ref{eq:nbh} is 
\begin{align}
\fF=\int_\Omega \big(\frac{1}{2} (\nabla^2 w)^T(\nabla^2 w)-q_0 w \big)d \Omega\notag.
\end{align}
With the aid of nonlocal Hessian operator $\tnabla^2 w$ and its variation $\tnabla^2 \tdelta w$, the residual and tangent stiffness matrix can be obtained with ease
\begin{align}
\mathbf R_g&=\tdelta \fF=\sum_{\Delta V_\bx\in \Omega} \Delta V_\bx\big((\tnabla^2 w)^T(\tnabla^2 \tdelta w)-q_0 \tdelta w\big)\notag\\
\mathbf K_g&=\tdelta^2 \fF=\sum_{\Delta V_\bx\in \Omega} \Delta V_\bx(\tnabla^2 \tdelta w)^T\tnabla^2 \tdelta w\notag.
\end{align}
The plate is discretized uniformly and the support radius is selected as $h=2.2\Delta \bx$. The weight function is $w(\br)=\frac{1}{r^2}$. The second-order hourglass control is exploited. The calculation of the nonlocal Hessian operator is given in \ref{app:h}. The deflection curves for several discretizations are compared with the analytic solution in Fig.\ref{fig:bhmC}. The contour of the deflection field for discretization of $20\times 20$ is shown in Fig.\ref{fig:bhmC}.
\begin{figure}
 \centering
 \subfigure[]{
 \label{fig:bhmW}
 \includegraphics[width=.4\textwidth]{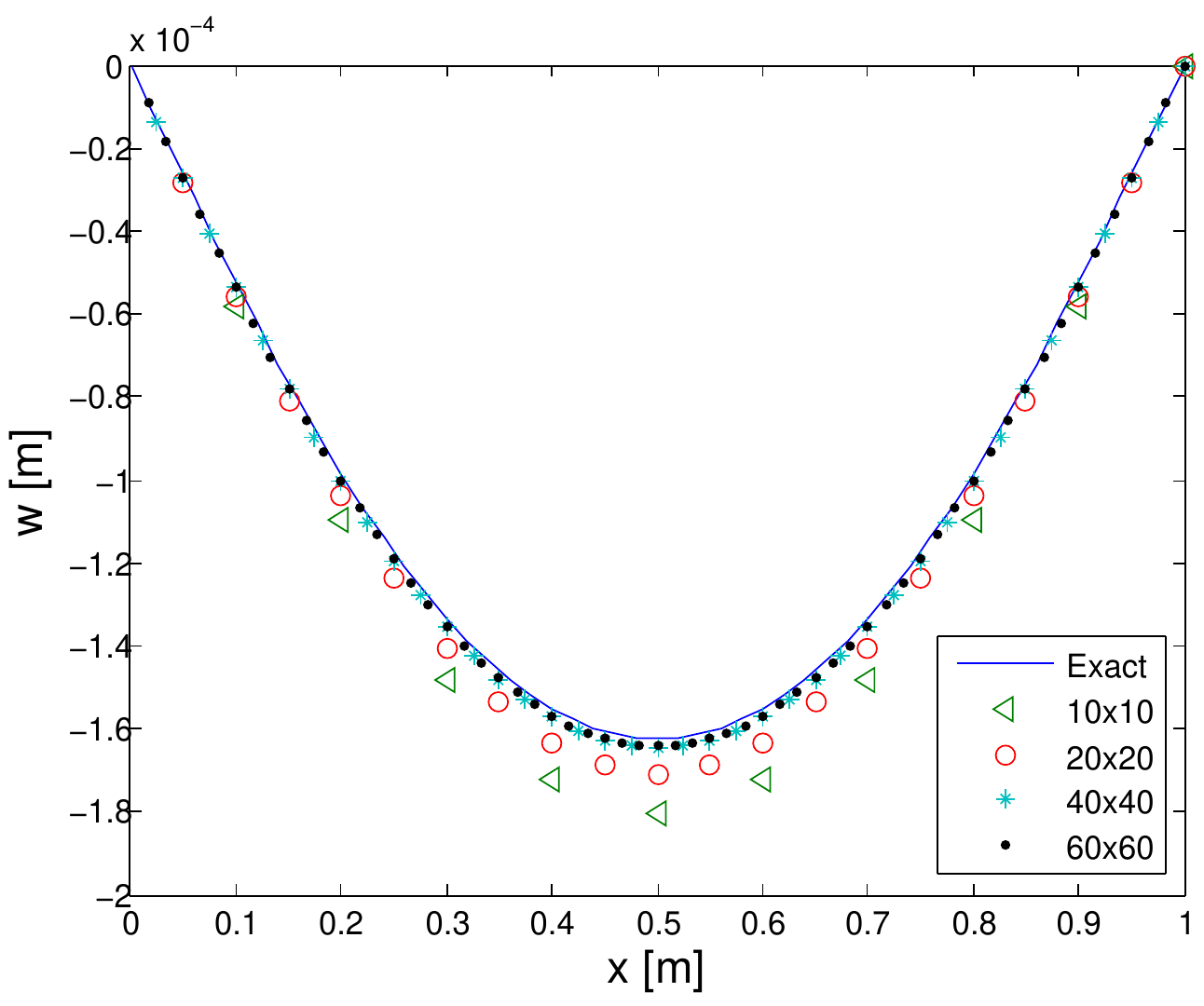}}
 \vspace{.1in}
 \subfigure[]{
 \label{fig:bhmC}
 \includegraphics[width=.4\textwidth]{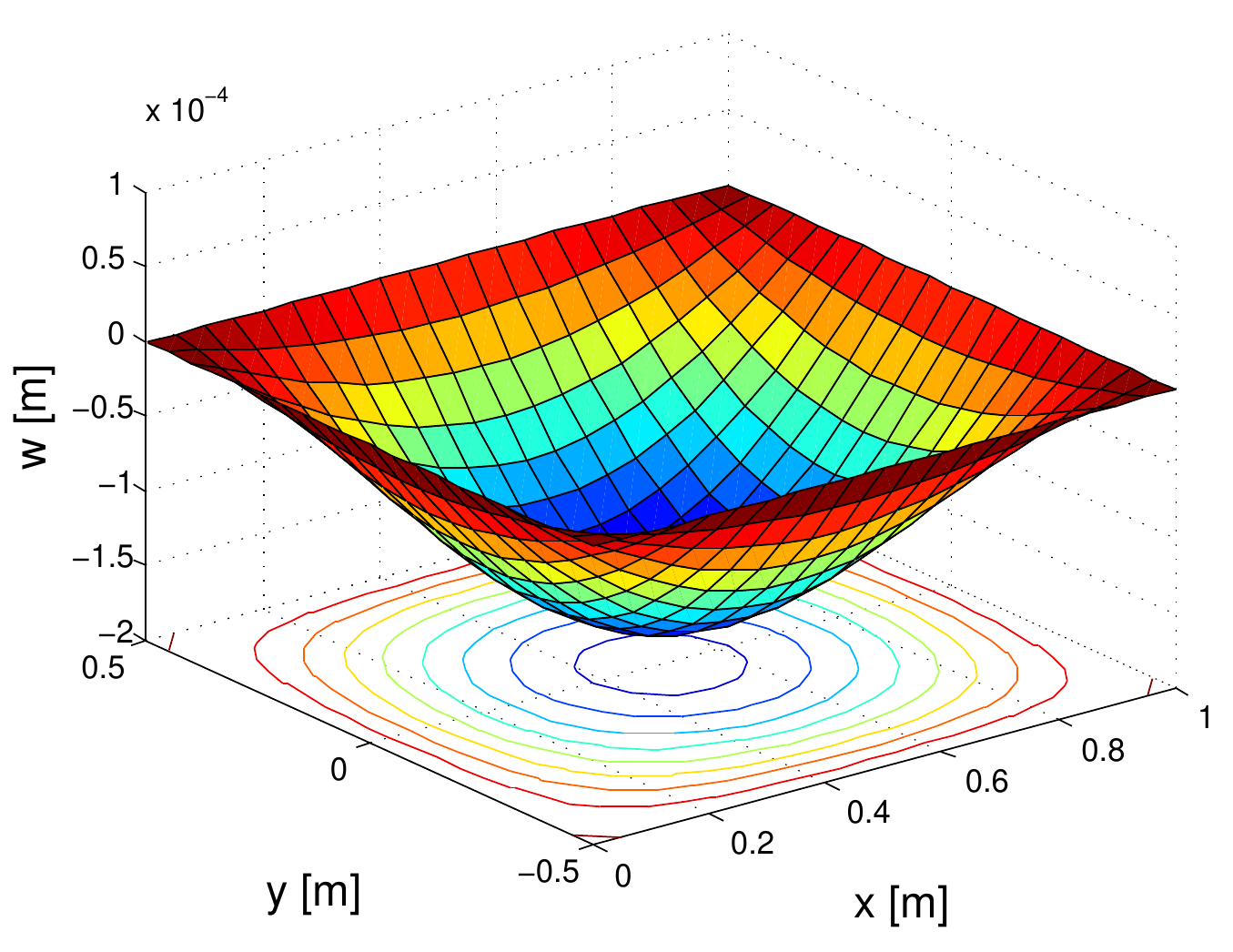}}\\
 \vspace{.3in}
\caption{(a) Deflection of section $y=0$ under different discretizations. (b) Contour of the deflection $w$ for discretization of 20x20.}
\label{fig:bhm}
\end{figure} 

\subsection{2D solid beam}
A two-dimensional cantilever beam loaded at the end with pure shear traction force is considered. The beam with dimensions of height of $D=12$m, length of $L=48$ m and shear load of parabola distribution. The analytical solution for the beam is \cite{timoshenko412theory,zhuang2010aspects}
\begin{align}
u_x&=\frac{P y}{6 EI}\big[(6 L-3 x)x+(2+\nu)(y^2-\frac{ \ud ^2}{4})\big]\label{eq:cantileverux}\\
u_y&=-\frac{P}{6 EI}\big[3 \nu y^2(L-x)+(4+5 \nu)\frac{ \ud ^2 x}{4}+(3L-x)x^2 \big]\label{eq:cantileveruy}\\
\sigma_{xx}(x,y)&=\frac{P(L-x)y}{I},\sigma_{yy}(x,y)=0,\tau_{xy}(x,y)=-\frac{P}{2 I}\big(\frac{ \ud ^2}{4}-y^2\big),\label{eq:cantileversigma}
\end{align}
where $(x,y)\in [0,L]\times [-D/2,D/2]$, $P=-1000$ N,$I=\frac{ \ud ^3}{12}$ and material parameters $E=10^8 $Pa,$\nu=0.3$. The particles on the left boundary are constrained by the exact displacements from Eq.\ref{eq:cantileverux} and Eq.\ref{eq:cantileveruy} and the loading on the right boundary follows Eq.\ref{eq:cantileversigma}.

Several discretizations with different particle grids $\Delta x\in \{D/5,D/10,D/20,D/30,D/60\}$ are tested. The displacement and stress for discretization $10\times 40$ and $20\times 80$ are shown in Fig.\ref{fig:beam2D10} and Fig.\ref{fig:beam2D20}, respectively. Good agreements are obtained between the numerical results and analytical results.
The L2-norm of displacement field with approximately convergent rate of $r=0.882$ is shown in Fig.\ref{fig:2dbeamL2}.


\begin{figure}
 \centering
 \subfigure[]{
 \label{fig:10Uy}
 \includegraphics[width=.3\textwidth]{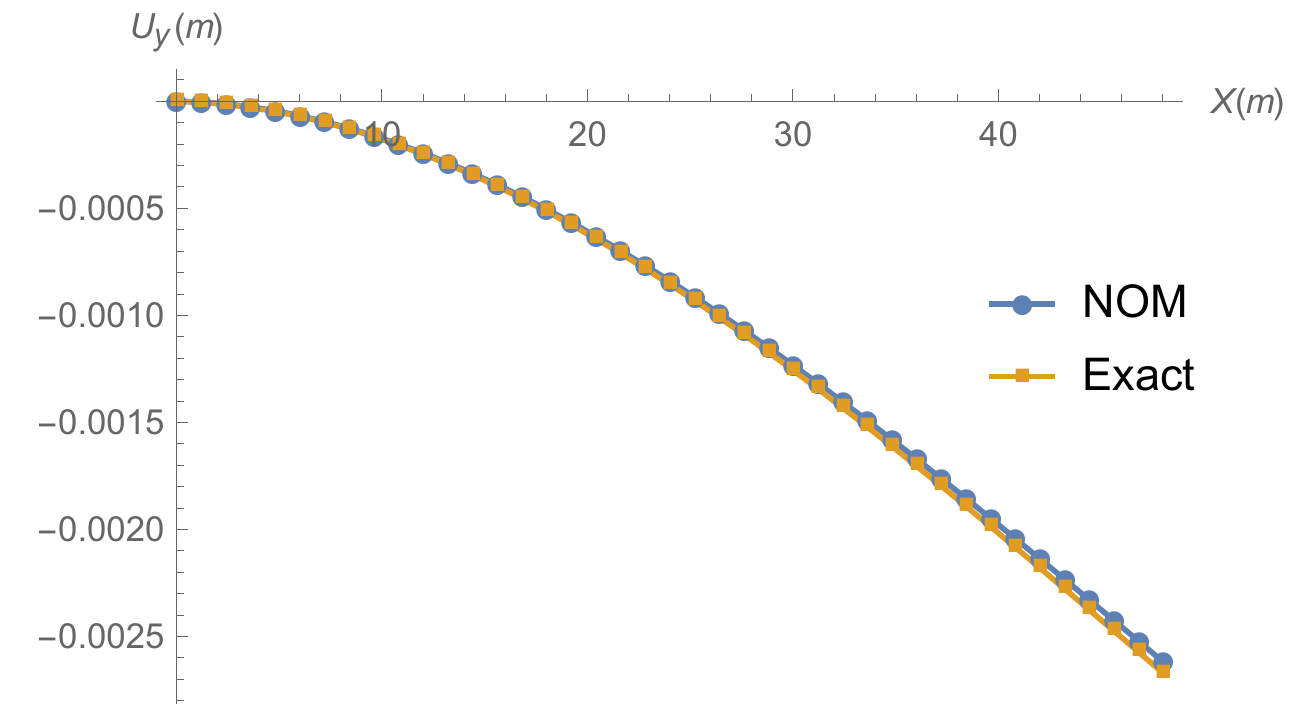}}
 \subfigure[]{
 \label{fig:10Ux}
 \includegraphics[width=.3\textwidth]{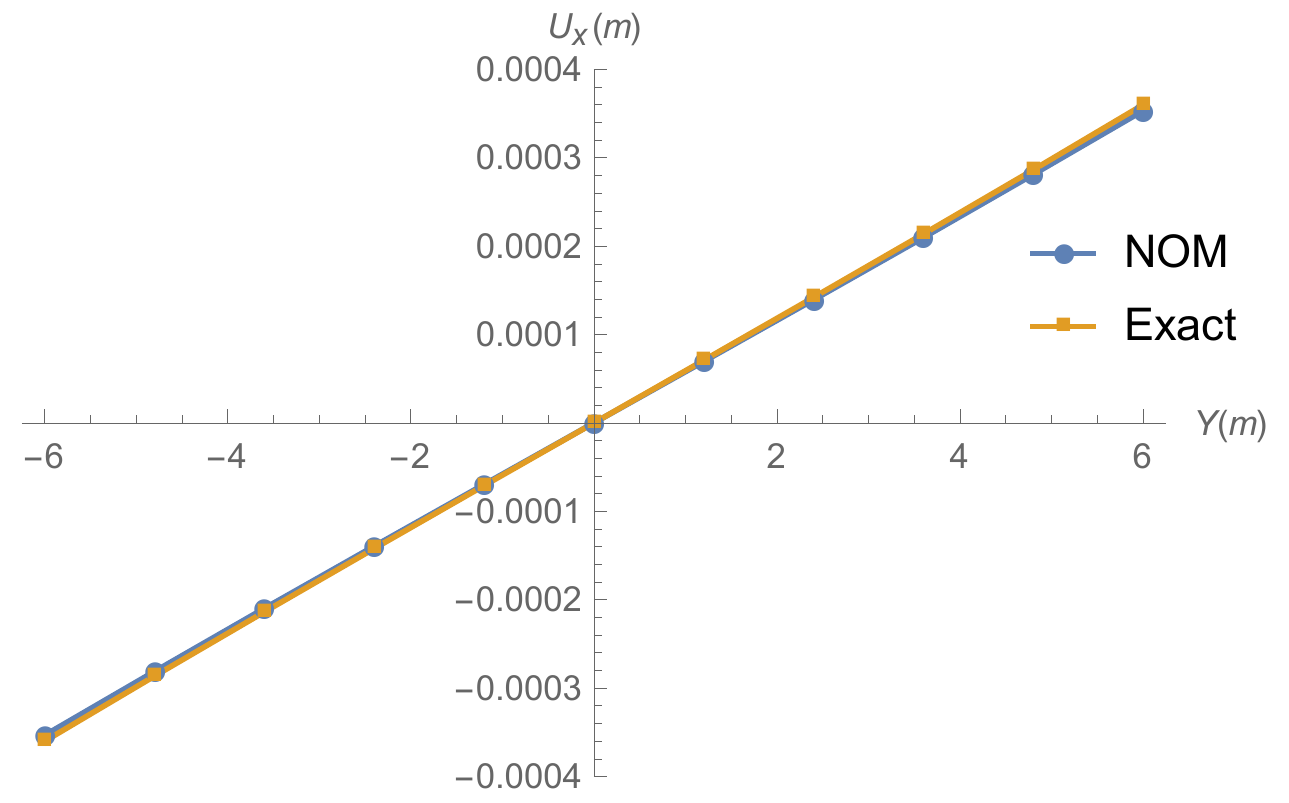}}
 \subfigure[]{
 \label{fig:10Sx}
 \includegraphics[width=.3\textwidth]{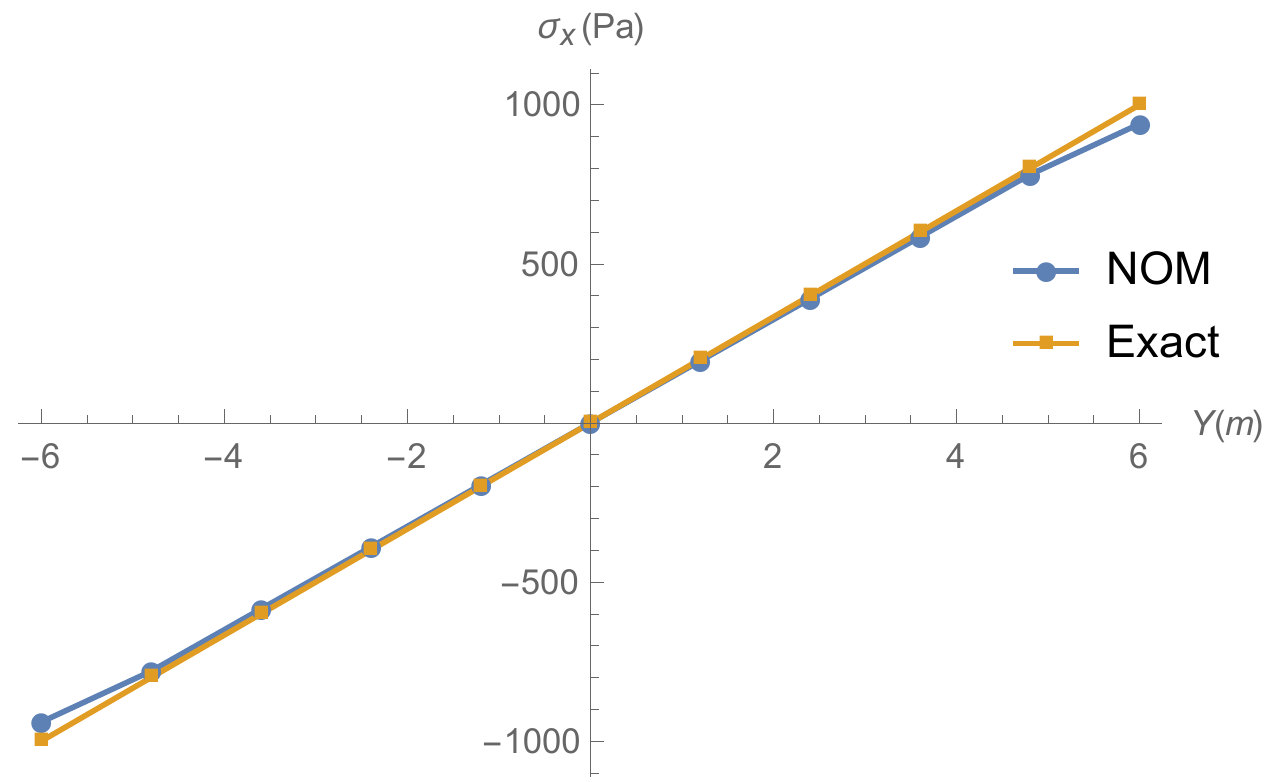}}
\caption{Beam under discretization $10\times 40$. (a) Displacement in $y$-direction for points on $x=0$; (b) displacement in $x$-direction for points on $y=L/2$; (c) stress in $x$-direction for points on $y=L/2$.}
\label{fig:beam2D10}
\end{figure}

\begin{figure}
 \centering
 \subfigure[]{
 \label{fig:20Uy}
 \includegraphics[width=.3\textwidth]{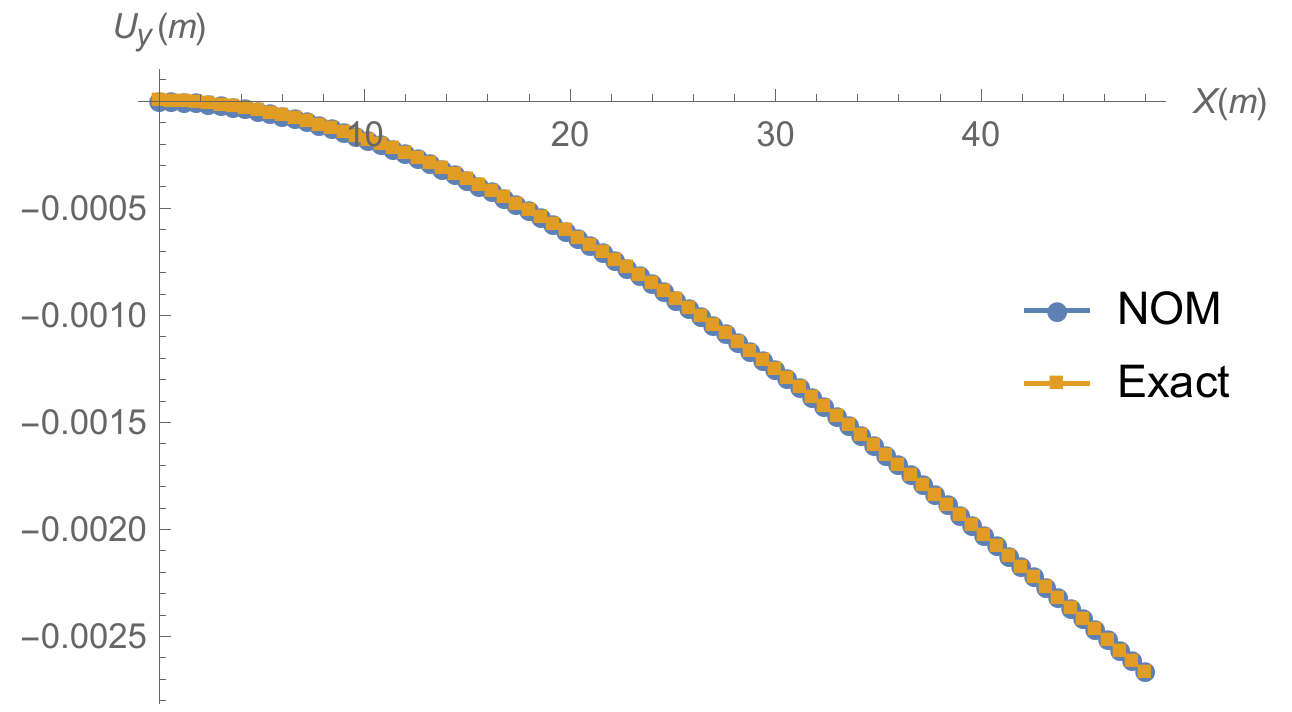}}
 \subfigure[]{
 \label{fig:20Ux}
 \includegraphics[width=.3\textwidth]{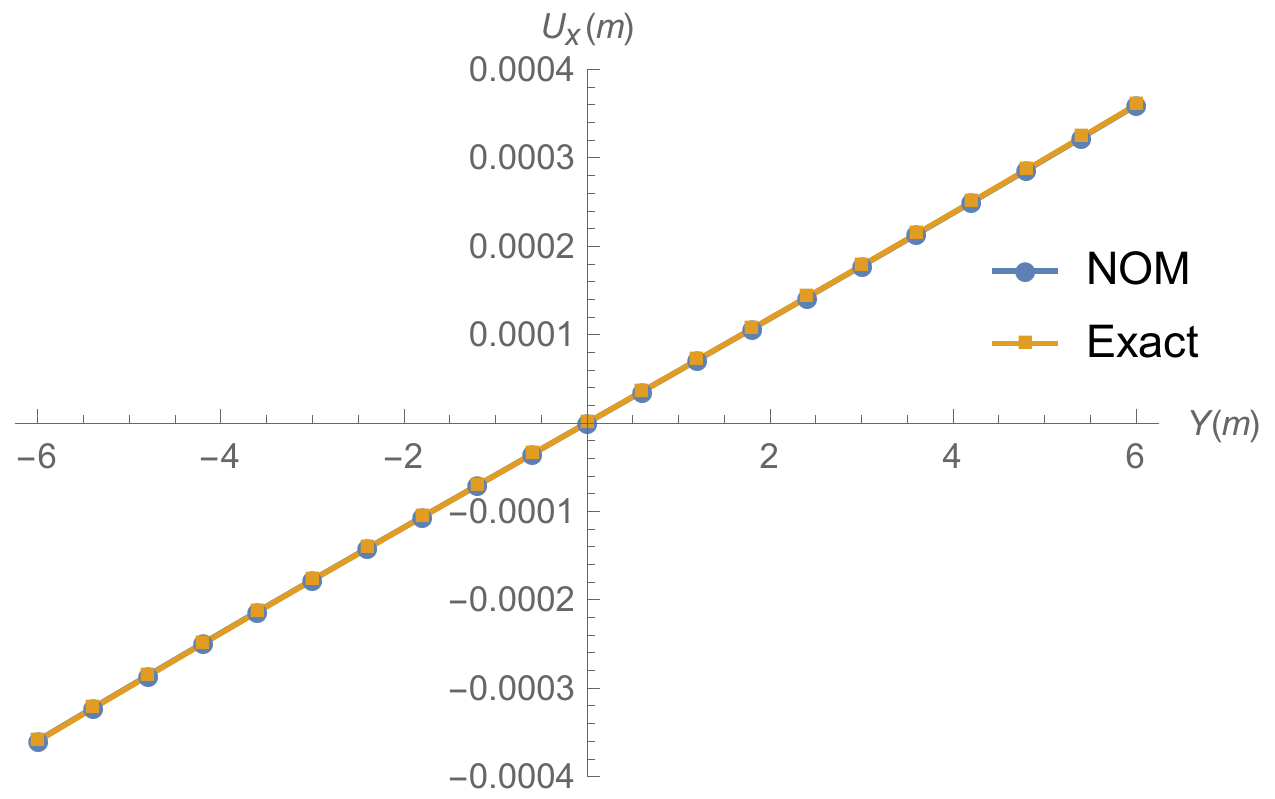}}
 \subfigure[]{
 \label{fig:20Sx}
 \includegraphics[width=.3\textwidth]{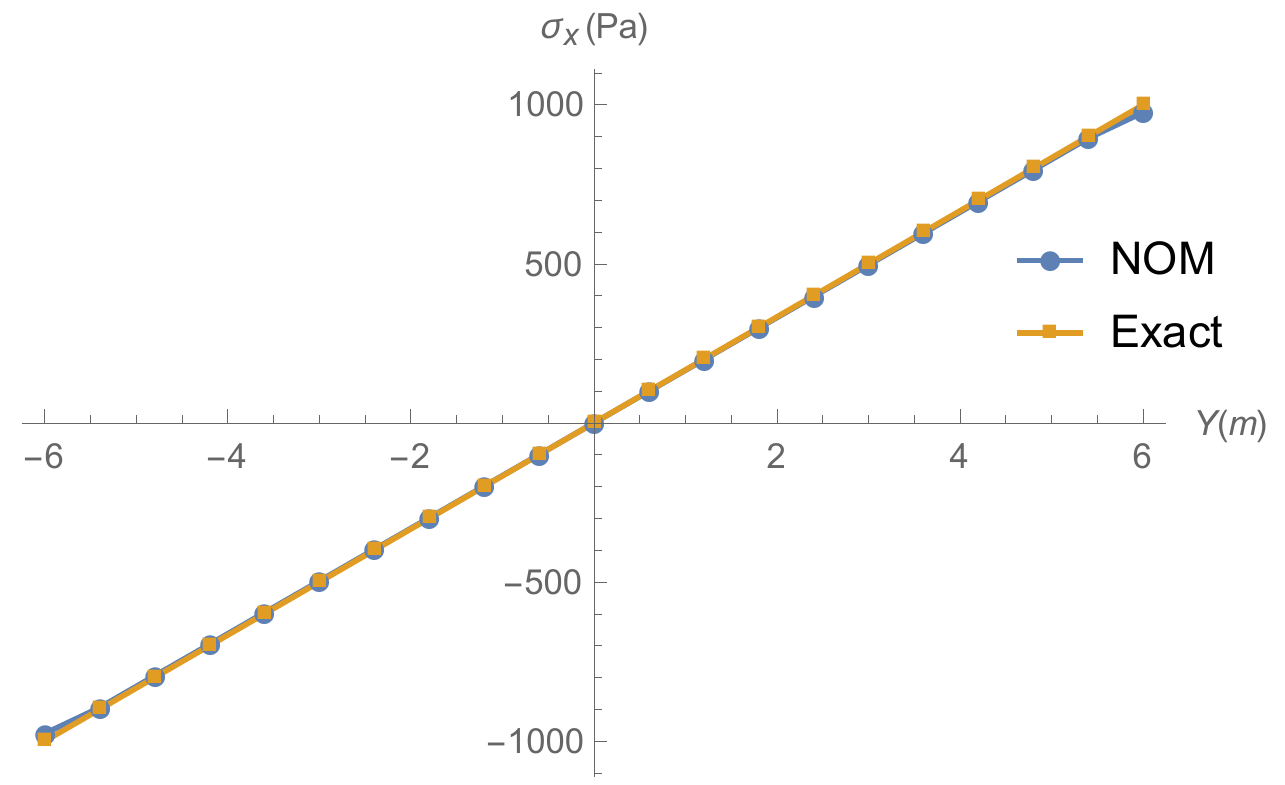}}
\caption{Beam under discretization $20\times 80$. (a) Displacement in $y$-direction for points on $x=0$; (b) displacement in $x$-direction for points on $y=L/2$; (c) stress in $x$-direction for points on $y=L/2$.}
\label{fig:beam2D20}
\end{figure}
\begin{figure}
	\centering
		\includegraphics[width=9cm]{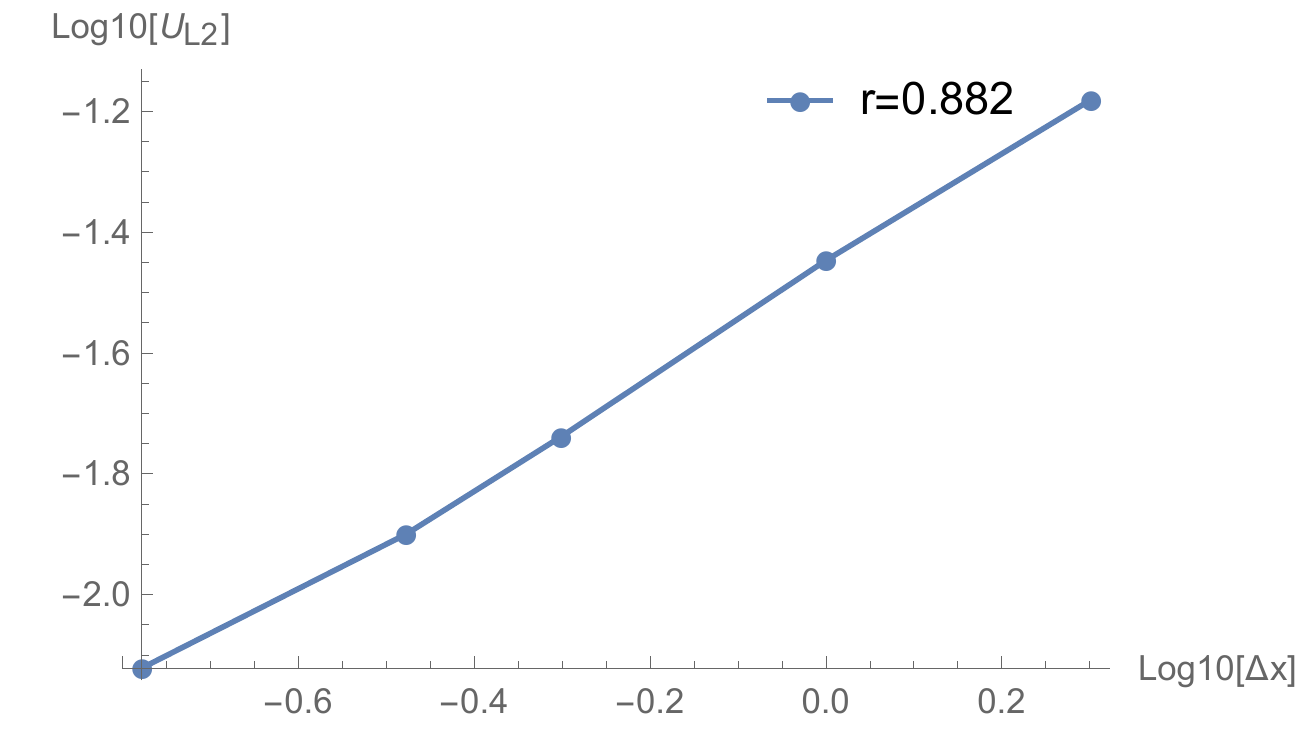}
	\caption{Convergence of displacement on $L_2$-norm.}
	\label{fig:2dbeamL2}
\end{figure}
\subsection{Plate with hole in tension}
This section solves the infinite plate with hole in tension and compares the numerical results by current method with that by analytical solutions. One quarter of the plate is modeled. For particles on $y=0$ ($x=0$) are fixed in $y$-direction ($x$-direction) by penalty method. The radius of the hole is $a=1$ and the length of the plate is $L=5$. 
\begin{figure}
 \centering
 \subfigure[]{
 \label{fig:hole1}
		\includegraphics[width=.47\textwidth]{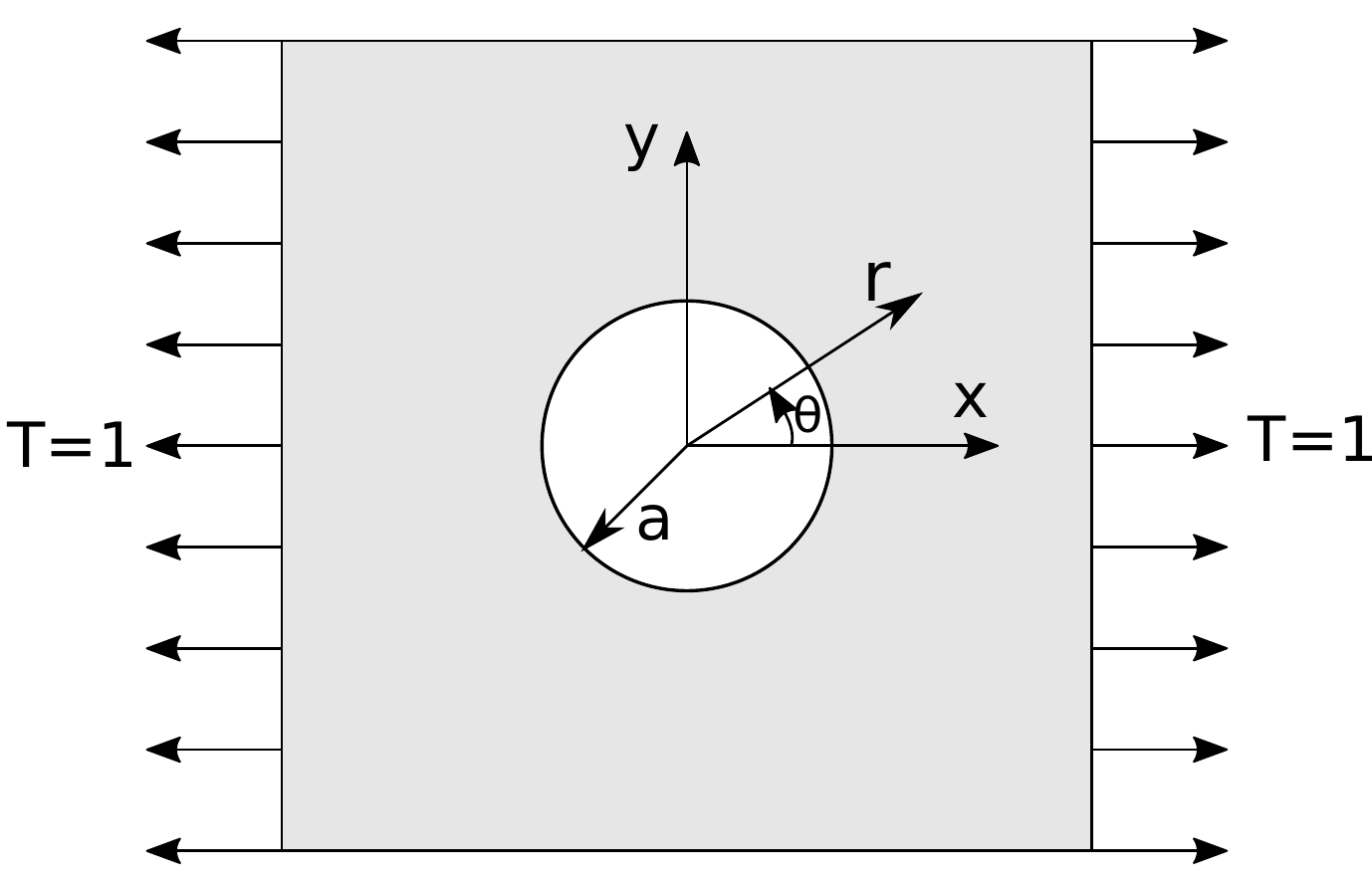}}
 \subfigure[]{
 \label{fig:hole2}
		\includegraphics[width=.3\textwidth]{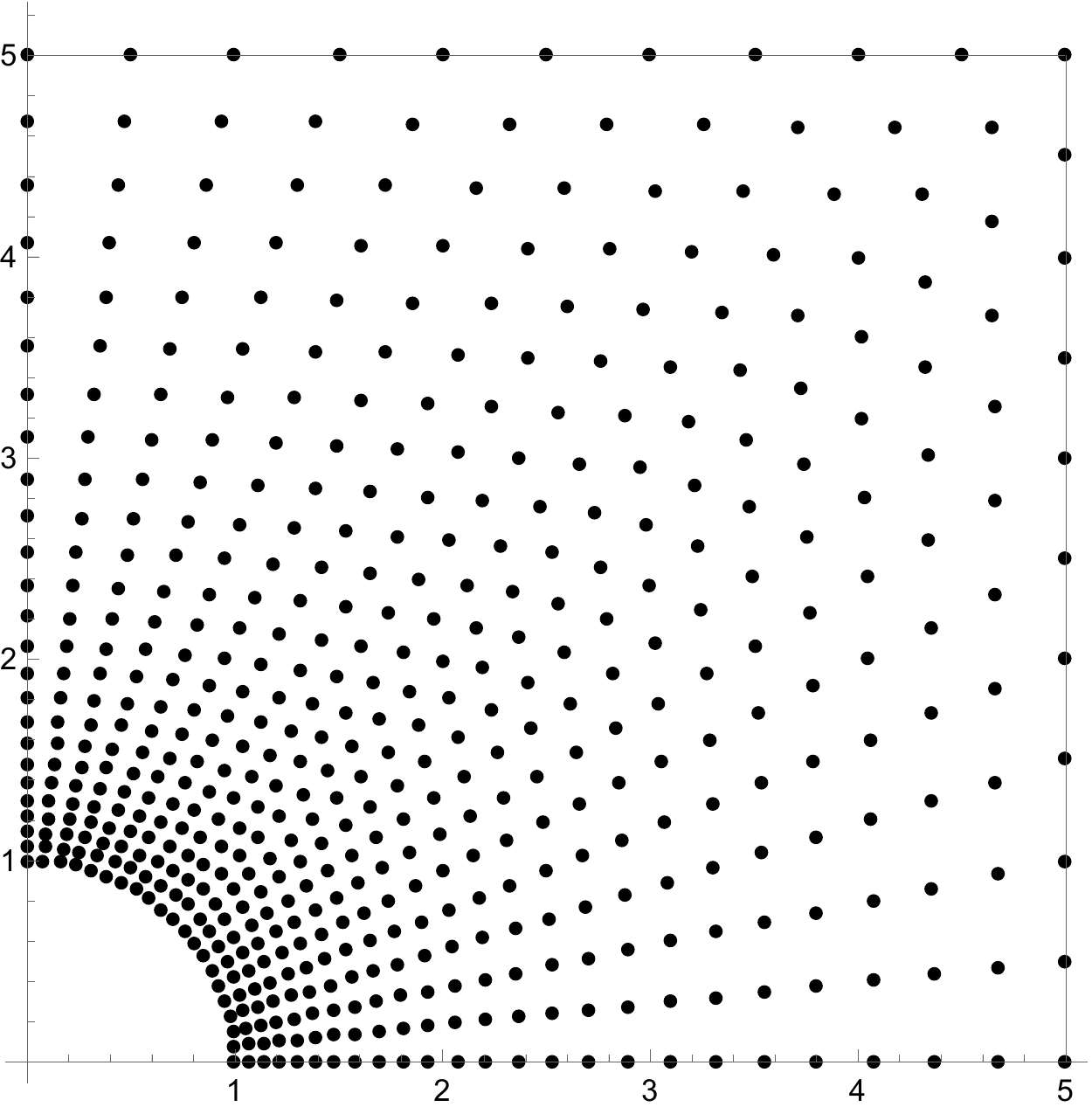} 
 }
\caption{(a) Setup of the plate with a hole; (b) discretization of the plate.}
\label{fig:hole3}
\end{figure}
The stresses in Cartesian coordinates \cite{boresi2010elasticity} are
\begin{align}
\sigma_{xx}(r,\theta )&=T-T{\frac {a^{2}}{r^{2}}}({\frac {3}{2}}\cos 2\theta +\cos 4\theta )+T{\frac {3a^{4}}{2r^{4}}}\cos 4\theta ,\notag\\
\sigma_{yy}(r,\theta )&=-T{\frac {a^{2}}{r^{2}}}({\frac {1}{2}}\cos 2\theta -\cos 4\theta )-T{\frac {3a^{4}}{2r^{4}}}\cos 4\theta,\label{eq:holeStressXY}\\
\tau_{xy}(r,\theta )&=-T{\frac {a^{2}}{r^{2}}}({\frac {1}{2}}\sin 2\theta +\sin 4\theta )+T{\frac {3a^{4}}{2r^{4}}}\sin 4\theta \notag
\end{align}

For plane stress conditions, the displacement can be expressed as
\begin{align}
u_{x}(r,\theta )&={\frac {Ta}{8\mu }}\left({\frac {r}{a}}(\kappa +1)\cos \theta +{\frac {2a}{r}}((1+\kappa )\cos \theta +\cos 3\theta )-{\frac {2a^{3}}{r^{3}}}\cos 3\theta \right),\notag
\\u_{y}(r,\theta )&={\frac {Ta}{8\mu }}\left({\frac {r}{a}}(\kappa -3)\sin \theta +{\frac {2a}{r}}((1-\kappa )\sin \theta +\sin 3\theta )-{\frac {2a^{3}}{r^{3}}}\sin 3\theta \right)
\end{align}
where $\mu=\frac {E}{2(1+\nu )}$, and $\kappa=\frac {3-\nu }{1+\nu }$.

For particles on $y=L (x=L)$ are applied with the surface traction force calculated by Eq.\ref{eq:holeStressXY}. The discretization with 11 nodes on left edge is shown in Fig.\ref{fig:hole2}. It should be noted that only the nodes are used and the area associated to nodes are constructed from the element area. The material parameters are $E=1000, \nu=0.3$. Three cases with total 525,2050,8019 nodes, respectively, are tested. 
The displacement and stress on polar coordinate $r=2 a$ are compared with the analytical solutions, as shown in Figs.(\ref{fig:40Ur},\ref{fig:40Ut},\ref{fig:40Stt}). The L2 norm of the displacement field by Eq.\ref{eq:uerror} are (0.0803,0.0371,0.0217) for three cases, respectively.
\begin{figure}
 \centering
 \subfigure[]{
 \label{fig:40Ur}
 \includegraphics[width=.3\textwidth]{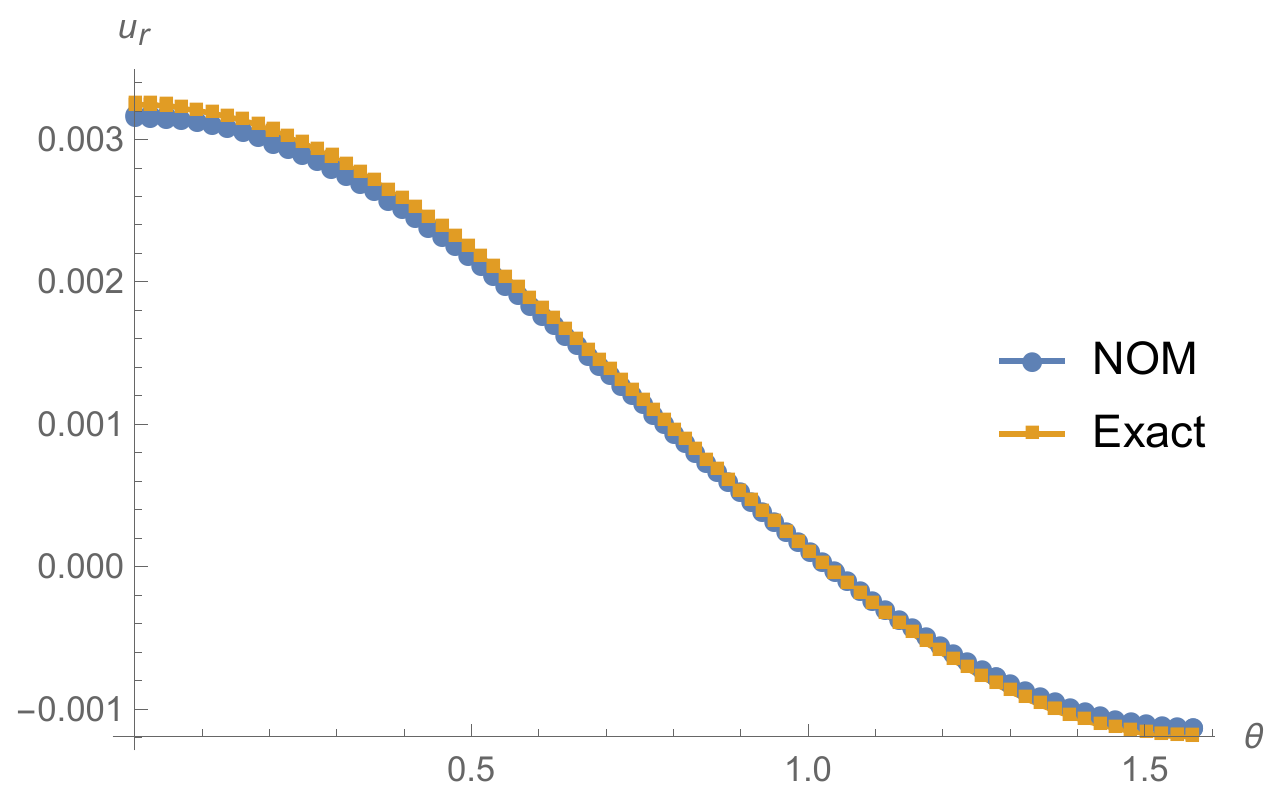}}
 \subfigure[]{
 \label{fig:40Ut}
 \includegraphics[width=.3\textwidth]{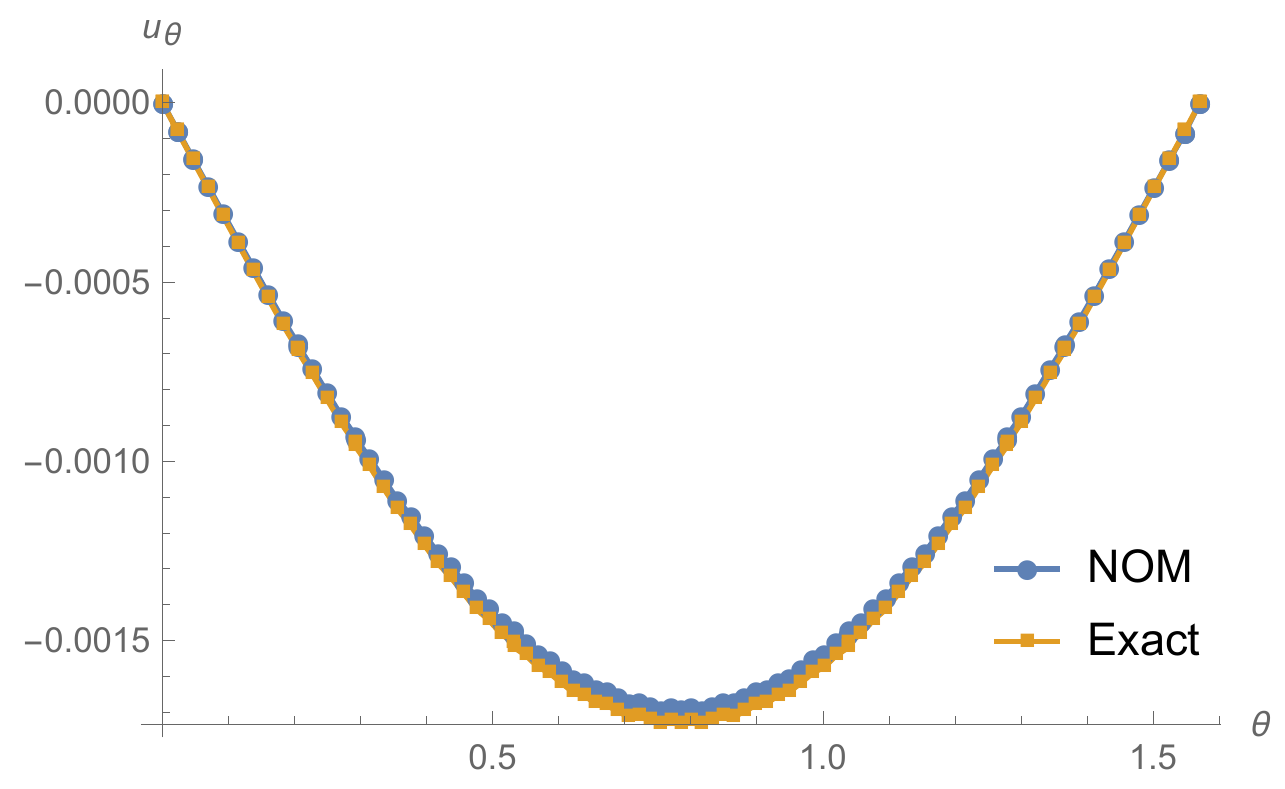}}
 \subfigure[]{
 \label{fig:40Stt}
 \includegraphics[width=.3\textwidth]{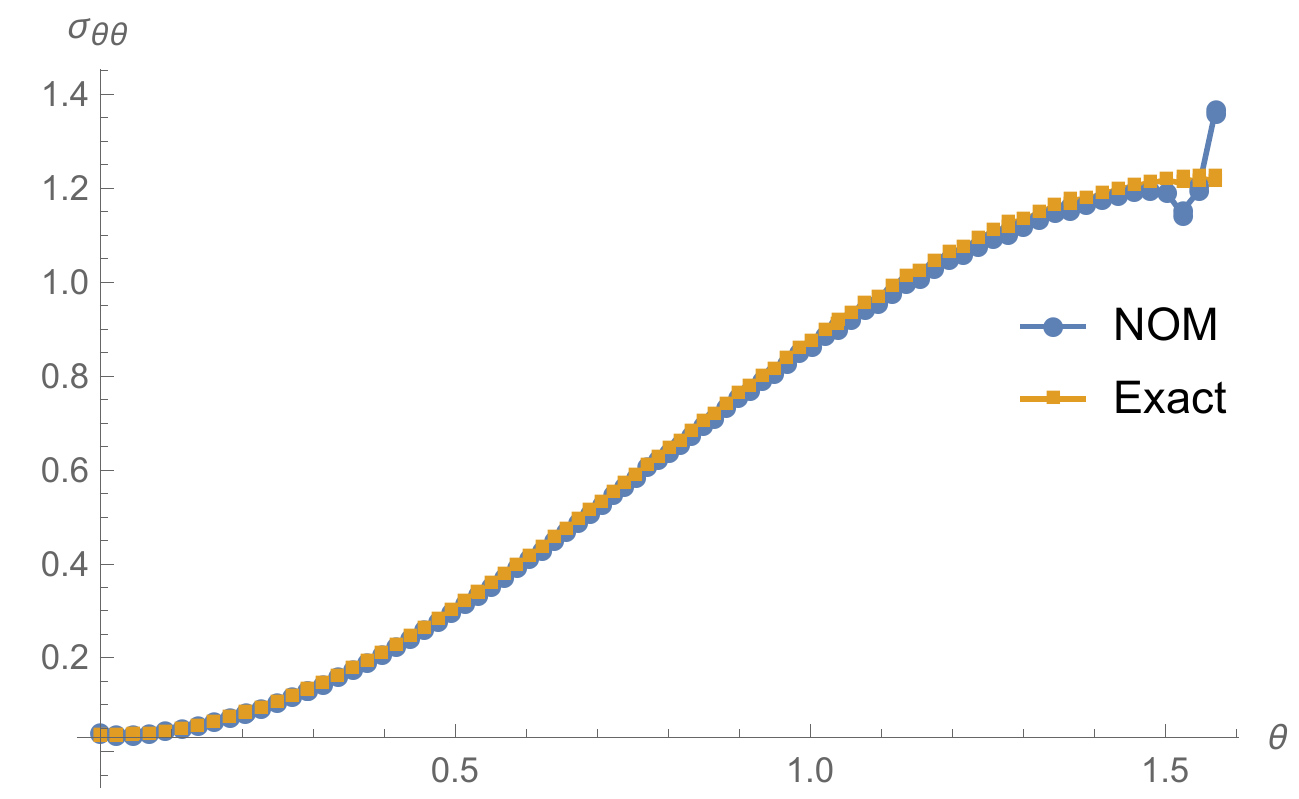}}
\caption{Exact results vs. numerical results. (a) $u_r$ for points on $r=2 a$; (b) $u_\theta$ for points on $r=2 a$; (c) $\sigma_{\theta\theta}$ for points on $r=2 a$.}
\label{fig:Hole2D40}
\end{figure}

\vspace{-6pt}
\section{Conclusions}\label{sec:end}
\vspace{-2pt}
We propose a nonlocal operator method for solving PDEs. The fundamental consituents in nonlocal operator method include the support, dual-support, nonlocal operators and hourglass energy functional. The support is the basis to define the nonlocal operators. The nonlocal operator is a generalization of the conventional differential operators. Under certain conditions such as support decreasing to an infinitesimal or linear field, the nonlocal operators converge to the local operators. On the other hand, the nonlocal operator is still valid in the case of field involving discontinuity since the nonlocal operator is defined by integral form. The dual-support as the dual concept of support allows the inhomogeneous discretization of the computational domain. The dual-support contributes to deriving the nonlocal strong discrete or continuous forms of different functionals by means of variational principles. 

The nonlocal operator is defined at one point but interacts with any other points in its support domain through nonlocal interactions. In this paper, the continuous form is solved by discretizing the computational domain into particles, and finally results in a discrete system based on nodal integration. Nodal integration method suffers the rank deficiency and hourglass mode (or zero-energy mode). In order to remove the hourglass mode,  the hourglass energy functional is proposed, which can suppress the hourglass modes in implicit/explicit analysis. 

The nonlocal operator method is consistent with the variational principle. The residual and tangent stiffness can be obtained with some matrix multiplication on common terms such as physical constitutions and nonlocal operators with variation. The nonlocal operator can be used to replace the traditional local operator of one-order or higher orders and thus obtains the discrete algebraic system of the PDEs with ease. In the example of nonlocal linear elasticity theory, the nonlocal operator method obtains the residual and tangent stiffness matrix concisely.

Several numerical examples include the deflection of cantilever beam and plate, the Poisson equation in 2D and eigenvalue problem, the problems in linear elastic mechanics are presented to illustrate the capabilities of the nonlocal operator method.
\vspace{-6pt}
\section*{Acknowledgments}
The authors acknowledge the supports from the COMBAT Program (Computational Modeling and Design of Lithium-ion Batteries, Grant No.615132) and NSFC (11772234), the Ministry of Science and Technology of China (SLDRCE14-B-31).
\vspace{-6pt}
\appendix

\section{Nonlocal Hessian operator in 1D,2D}\label{app:h}
In the case of 1 dimension, all operators are scalar-type. The second derivative and its variation in 1D can be written as
\begin{align}
\frac{ \ud ^2 u}{ \ud x^2} &=2\int_{\cS} w(r) u_{r}(r^2- \frac{K_3}{K_2} r) \ud V \cdot K_4^{-1}\\
\tnabla^2 \tdelta u &=2\int_{\cS} w(r) (\tdelta u'-\tdelta u)(r^2- \frac{K_3}{K_2} r) \ud V \cdot K_4^{-1}. \label{eq:avsd}
\end{align}

For simplicity, we consider Hessian operator in two dimensions, i.e. $\br=(x,y)^T$. 
The 2-order shape tensor is
\begin{align}
\bK_2=\sum_{\cS} w(\br)\Delta V'\begin{bmatrix}
 x^2 & x y\\
 x y & y^2\\
\end{bmatrix}.
\end{align}
The 3-order shape tensor is
\begin{align}
\bK_3=(\bK_3^x,\bK_3^y)=\Big(\sum_{\cS} w(\br)\Delta V' x\begin{bmatrix}
 x^2 & x y\\
 x y & y^2\\
\end{bmatrix},&\sum_{\cS} w(\br)\Delta V' y\begin{bmatrix}
 x^2 & x y\\
 x y & y^2\\
\end{bmatrix}\Big).
\end{align}
Therefore, the calculation of $\bK_3 \bK_2^{-1} \br$ is 
\begin{align}
\bK_3 \bK_2^{-1} \br=(\bK_3^x \bK_2^{-1} \br,\bK_3^y \bK_2^{-1} \br).
\end{align}
The 4-order shape tensor is
\begin{align}
\bK_4=\begin{bmatrix}\bK_4^{xx}&\bK_4^{xy}\\\bK_4^{yx}&\bK_4^{yy}\end{bmatrix}=\sum_{\cS} w(\br)\Delta V'\begin{bmatrix} x^2\begin{bmatrix}
 x^2 & x y\\
 x y & y^2\\
\end{bmatrix}& x y\begin{bmatrix}
 x^2 & x y\\
 x y & y^2\\
\end{bmatrix}\\
yx\begin{bmatrix}
 x^2 & x y\\
 x y & y^2\\
\end{bmatrix}& y^2\begin{bmatrix}
 x^2 & x y\\
 x y & y^2\\
\end{bmatrix}
\end{bmatrix}.
\end{align}
It should be noted that the rank of $\bK_4$ is 3 since there are only three independent variables in the 2D nonlocal Hessian operator where ${\tfrac{\partial^2 \delta u}{\partial x\partial y}=\tfrac{\partial^2 \delta u}{\partial y\partial x}}$. It is convenient to write ${\nabla^2 \delta u=(\tfrac{\partial^2 \delta u}{\partial x\partial x},\tfrac{\partial^2 \delta u}{\partial x\partial y},\tfrac{\partial^2 \delta u}{\partial y\partial y})}$. 
\vspace{-6pt}
\section*{References}
\vspace{-6pt}
\bibliographystyle{unsrt}
\bibliography{horizonElement/pdemim}

\begin{thebibliography}{10}

\bibitem{zienkiewicz1977finite}
Olgierd~Cecil Zienkiewicz, Robert~Leroy Taylor, Olgierd~Cecil Zienkiewicz, and
  Robert~Lee Taylor.
\newblock {\em The finite element method}, volume~36.
\newblock McGraw-hill London, 1977.

\bibitem{lucy1977numerical}
Leon~B Lucy.
\newblock A numerical approach to the testing of the fission hypothesis.
\newblock {\em The astronomical journal}, 82:1013--1024, 1977.

\bibitem{nayroles1992generalizing}
B~Nayroles, G~Touzot, and P~Villon.
\newblock Generalizing the finite element method: diffuse approximation and
  diffuse elements.
\newblock {\em Computational mechanics}, 10(5):307--318, 1992.

\bibitem{Belytschko1994}
Ted Belytschko, Yun~Yun Lu, and Lei Gu.
\newblock Element-free galerkin methods.
\newblock {\em International journal for numerical methods in engineering},
  37(2):229--256, 1994.

\bibitem{LiuJunZhang1995}
Wing~Kam Liu, Sukky Jun, and Yi~Fei Zhang.
\newblock Reproducing kernel particle methods.
\newblock {\em International journal for numerical methods in fluids},
  20(8-9):1081--1106, 1995.

\bibitem{babuvska1997partition}
Ivo Babu{\v{s}}ka and Jens~M Melenk.
\newblock The partition of unity method.
\newblock {\em International journal for numerical methods in engineering},
  40(4):727--758, 1997.

\bibitem{duarte2000generalized}
C~Armando Duarte, Ivo Babu{\v{s}}ka, and J~Tinsley Oden.
\newblock Generalized finite element methods for three-dimensional structural
  mechanics problems.
\newblock {\em Computers \& Structures}, 77(2):215--232, 2000.

\bibitem{duarte1996hp}
C~Armando Duarte and J~Tinsley Oden.
\newblock An hp adaptive method using clouds.
\newblock {\em Computer methods in applied mechanics and engineering},
  139(1-4):237--262, 1996.

\bibitem{onate1996finite}
E~Onate, S~Idelsohn, OC~Zienkiewicz, and RL~Taylor.
\newblock A finite point method in computational mechanics. applications to
  convective transport and fluid flow.
\newblock {\em International journal for numerical methods in engineering},
  39(22):3839--3866, 1996.

\bibitem{liszka1984interpolation}
Tadeusz Liszka.
\newblock An interpolation method for an irregular net of nodes.
\newblock {\em International Journal for Numerical Methods in Engineering},
  20(9):1599--1612, 1984.

\bibitem{aluru2000point}
NR~Aluru.
\newblock A point collocation method based on reproducing kernel
  approximations.
\newblock {\em International Journal for Numerical Methods in Engineering},
  47(6):1083--1121, 2000.

\bibitem{hu2011error}
Hsin-Yun Hu, Jiun-Shyan Chen, and Wei Hu.
\newblock Error analysis of collocation method based on reproducing kernel
  approximation.
\newblock {\em Numerical Methods for Partial Differential Equations},
  27(3):554--580, 2011.

\bibitem{Silling2007}
Stewart~A Silling, M~Epton, O~Weckner, J~Xu, and E~Askari.
\newblock Peridynamic states and constitutive modeling.
\newblock {\em Journal of Elasticity}, 88(2):151--184, 2007.

\bibitem{Ren2015}
HL~Ren, XY~Zhuang, YC~Cai, and T~Rabczuk.
\newblock Dual-horizon peridynamics.
\newblock {\em International Journal for Numerical Methods in Engineering},
  2016.

\bibitem{gingold1977smoothed}
Robert~A Gingold and Joseph~J Monaghan.
\newblock Smoothed particle hydrodynamics: theory and application to
  non-spherical stars.
\newblock {\em Monthly notices of the royal astronomical society},
  181(3):375--389, 1977.

\bibitem{nguyen2008meshless}
Vinh~Phu Nguyen, Timon Rabczuk, St{\'e}phane Bordas, and Marc Duflot.
\newblock Meshless methods: a review and computer implementation aspects.
\newblock {\em Mathematics and computers in simulation}, 79(3):763--813, 2008.

\bibitem{chen2017meshfree}
Jiun-Shyan Chen, Michael Hillman, and Sheng-Wei Chi.
\newblock Meshfree methods: progress made after 20 years.
\newblock {\em Journal of Engineering Mechanics}, 143(4):04017001, 2017.

\bibitem{Belytschko19}
T.~Belytschko, N.~Mo{\"e}s, S.~Usui, and C.~Parimi.
\newblock Arbitrary discontinuities in finite elements.
\newblock {\em International Journal for Numerical Methods in Engineering},
  50(4):993--1013, 2001.

\bibitem{eringen2002nonlocal}
A~Cemal Eringen.
\newblock {\em Nonlocal continuum field theories}.
\newblock Springer Science \& Business Media, 2002.

\bibitem{Silling2000}
S.A. Silling.
\newblock Reformulation of elasticity theory for discontinuities and long-range
  forces.
\newblock {\em Journal of the Mechanics and Physics of Solids}, 48(1):175--209,
  2000.

\bibitem{bavzant2002nonlocal}
Zden{\v{e}}k~P Ba{\v{z}}ant and Milan Jir{\'a}sek.
\newblock Nonlocal integral formulations of plasticity and damage: survey of
  progress.
\newblock {\em Journal of Engineering Mechanics}, 128(11):1119--1149, 2002.

\bibitem{gunzburger2010nonlocal}
Max Gunzburger and Richard~B Lehoucq.
\newblock A nonlocal vector calculus with application to nonlocal boundary
  value problems.
\newblock {\em Multiscale Modeling \& Simulation}, 8(5):1581--1598, 2010.

\bibitem{du2013nonlocal}
Qiang Du, Max Gunzburger, Richard~B Lehoucq, and Kun Zhou.
\newblock A nonlocal vector calculus, nonlocal volume-constrained problems, and
  nonlocal balance laws.
\newblock {\em Mathematical Models and Methods in Applied Sciences},
  23(03):493--540, 2013.

\bibitem{ren2017dual}
HL~Ren, XY~Zhuang, and T~Rabczuk.
\newblock Dual-horizon peridynamics: A stable solution to varying horizons.
\newblock {\em Computer Methods in Applied Mechanics and Engineering},
  318:762--782, 2017.

\bibitem{polizzotto2001nonlocal}
Castrenze Polizzotto.
\newblock Nonlocal elasticity and related variational principles.
\newblock {\em International Journal of Solids and Structures},
  38(42-43):7359--7380, 2001.

\bibitem{timoshenko1959theory}
Stephen~P Timoshenko and Sergius Woinowsky-Krieger.
\newblock {\em Theory of plates and shells}.
\newblock McGraw-hill, 1959.

\bibitem{timoshenko412theory}
S~Timoshenko and JN~Goodier.
\newblock {\em Theory of elasticity (3rd edn).}, volume 412.
\newblock McGraw-Hill: New York, 1970.

\bibitem{zhuang2010aspects}
Xiaoying Zhuang and Charles Augarde.
\newblock Aspects of the use of orthogonal basis functions in the element-free
  galerkin method.
\newblock {\em International Journal for Numerical Methods in Engineering},
  81(3):366--380, 2010.

\bibitem{boresi2010elasticity}
Arthur~P Boresi, Ken Chong, and James~D Lee.
\newblock {\em Elasticity in engineering mechanics}.
\newblock John Wiley \& Sons, 2010.

\end{thebibliography}
\end{document}